\documentclass[%
superscriptaddress,
 amsmath,amssymb,
 aps,
 pre,
 twocolumn,
]{revtex4-2}

\def\s{\sigma}

\usepackage[utf8]{inputenc} 
\usepackage[T1]{fontenc}    
\usepackage{hyperref}       
\usepackage{url}            
\usepackage{booktabs}       
\usepackage{amsfonts}       
\usepackage{nicefrac}       
\usepackage{microtype}      
\usepackage{xcolor}         
\usepackage{amsmath,amsfonts,amssymb}
\usepackage{color}
\usepackage{graphicx}
\usepackage{lmodern}
\usepackage{float}

\begin{document}

\title{Machine-learning-assisted Monte Carlo \\ fails at sampling computationally hard problems}

\author{Simone Ciarella}
\thanks{These authors contributed equally. Email: simone.ciarella@ens.fr, jeanne.trinquier@ens.fr}
 \affiliation{Laboratoire de Physique de l'Ecole Normale Sup\'erieure, ENS, Universit\'e PSL, CNRS, Sorbonne Universit\'e, Universit\'e de Paris, F-75005 Paris, France
}

\author{Jeanne Trinquier}
\thanks{These authors contributed equally. Email: simone.ciarella@ens.fr, jeanne.trinquier@ens.fr}
\affiliation{
Sorbonne Universit\'e, CNRS, Institut de Biologie Paris Seine, Biologie
Computationnelle et Quantitative LCQB, F-75005 Paris, France
}
 \affiliation{Laboratoire de Physique de l'Ecole Normale Sup\'erieure, ENS, Universit\'e PSL, CNRS, Sorbonne Universit\'e, Universit\'e de Paris, F-75005 Paris, France
}

\author{Martin Weigt}
\affiliation{
Sorbonne Universit\'e, CNRS, Institut de Biologie Paris Seine, Biologie
Computationnelle et Quantitative LCQB, F-75005 Paris, France
}

\author{Francesco Zamponi}
 \affiliation{Laboratoire de Physique de l'Ecole Normale Sup\'erieure, ENS, Universit\'e PSL, CNRS, Sorbonne Universit\'e, Universit\'e de Paris, F-75005 Paris, France
}

\date{\today}

\begin{abstract}
Several strategies have been recently proposed in order to improve Monte Carlo sampling efficiency using machine learning tools. Here, we challenge these methods by considering a class of problems that are known to be 
exponentially hard to sample using conventional local Monte Carlo at low enough temperatures. In particular, we study the antiferromagnetic Potts model on a random graph, which reduces to the coloring of random graphs at zero temperature. We test several machine-learning-assisted Monte Carlo approaches, and we find that they all fail. Our work thus provides good benchmarks for future proposals for smart sampling algorithms.
\end{abstract}

\maketitle

\section{Introduction}
\label{sec:intro}

\subsection{Motivations}

Sampling from a given {\it target} probability distribution $P_t(\s_1,\cdots,\s_N)$ over $N$ degrees of freedom can become extremely hard when $N$ is large.
A universal (i.e. system-independent) strategy for sampling consists in starting from a random configuration of $\s = \{\s_i\}_{i=1,\cdots,N}$, and generating a {\it local} Monte Carlo Markov Chain (MCMC), by sequentially proposing an update of one of the $\s_i$, and accepting or rejecting it with a proper probability (e.g. Metropolis-Hastings), until convergence~\cite{Krauth}. 
However, for large $N$, the convergence time of the MCMC can grow exponentially in $N$, because of non-trivial long-range correlations that make local decorrelation extremely hard~\cite{MS06}.

A solution to this problem consists in identifying the proper set of correlated variables, and proposing {\it global} updates of such variables together, in such a way to speed up convergence~\cite{swendsen1992new}. However, this process is not universal, because it relies on the proper identification of system-dependent correlations, which is not always possible. For instance, in disordered systems such as spin glasses, the nature of correlated domains is extremely elusive and proper global moves are not easy to identify~\cite{Jorg2005,Zhu2015}. Another approach, which has been particularly successful in atomistic models of glasses, consists in unconstraining some degrees of freedom, evolve them and constrain them back~\cite{NBC17,kapteijns2019fast,Ciarella2021a,hagh2022transient,ozawa22}, but again it is model-specific.
Alternative proposals based on a renormalization group approach~\cite{Li2018,marchand2022wavelet} also rely on the identification of system-dependent collective variables.

A recently developed line of research, see e.g.~\cite{Wu2019,PhysRevE.101.053312,gabrie2021adaptive,wu21,Hibat-Allah2021,fan2021finding,schuetz2022graph,condmat7020038}, proposed to solve the problem in an elegant and universal way, by {\it machine learning} proper MCMC moves. In a nutshell, the idea is to learn an auxiliary probability distribution $P_a(\s)$, which (i) can be sampled efficiently (e.g. linearly in $N$) and (ii) provides a good approximation of the target probability. Then, the hope is to use the auxiliary distribution to propose smart MCMC moves. Using this strategy with autoregressive architectures that ensure efficient sampling, some authors found convergence speedup~\cite{Wu2019,PhysRevE.101.053312,wu21}, but others found less promising results~\cite{condmat7020038}.

In order to make these studies more systematic, and really assess the performance of the method, it is important to have good benchmarks, i.e. problems that are guaranteed to be {\it really} hard to sample by local MCMC.
In the early 90s, the very same problem had to be faced to assess the performance of local search algorithms that looked for solution of optimization or satisfiability problems~\cite{cheeseman1991really}.
In that case, the problem of generating good benchmarks was solved by introducing an ensemble of {\it random instances} of the problem under study~\cite{cheeseman1991really,kirkpatrick1994critical,selman1996generating,monasson1999determining}. It was later shown, both numerically and analytically, that these random optimization/satisfiability problems require a time scaling exponentially in $N$ for proper sampling at low enough temperatures in certain regions of parameter space~\cite{MS06}.
Hence, they provide very good benchmarks for sampling algorithms.
Yet, the recent attempts to apply machine learning methods to speed-up sampling have not considered these benchmarks.

In this paper, we consider a prototypical hard-to-sample random problem, namely the coloring of random graphs, and we show that all the proposed methods fail to solve it.
Our results confirm that this class of problems are a real challenge for sampling methods, even assisted by smart  machine-learned moves. The model investigated in~\cite{condmat7020038}
possibly belong to this class.
In addition, we discuss some practical issues such as {\it mode-collapse} in learning the auxiliary model, which happens when the target probability distribution has multiple peaks and the auxiliary model only learns one (or a subset) of them.

\subsection{State of the art}

Before proceeding, we provide a short review of the papers that motivated our study. Because the field is evolving rapidly, this does not aim at being an exhaustive review, and despite our best efforts, it is possible that we missed some relevant references.

Ref.~\cite{merchan2016sufficiency} considered the general problem of whether a target probability distribution $P_t$ can be approximated by a simpler one $P_a$, in particular by considering the Kullback-Leibler (KL) divergence
\begin{equation}
 D_{KL}(P_t||P_a) = \left\langle \log\frac{P_t(\s)}{P_a(\s)} \right\rangle_{P_t} \ .
\end{equation}
If this quantity is proportional to $N$ for $N\to\infty$, then $P_t(\s)/P_a(\s)$ is typically exponential in $N$, and as a result samples proposed from $P_a$ are very unlikely to be accepted in $P_t$. A small $D_{KL}(P_t||P_a)/N$ (ideally vanishing for $N\to\infty$) seems therefore to be a necessary condition for a good auxiliary probability, which provides a quantitative measure of condition~(ii) above. 
Ref.~\cite{merchan2016sufficiency} suggested, by using small disordered systems (${N\sim20}$), that there might be a phase transition, for $N\to\infty$, separating a phase where $D_{KL}(P_t||P_a)/N$ vanishes identically and a phase where it is positive. 

Ref.~\cite{Wu2019} proposed, more specifically, to use autoregressive models as tractable architectures for $P_a$. In these architectures, $P_a$ is represented using Bayes' rule,
\begin{equation}\label{eq:BayesAR}
    P_a(\s) = P_a^1(\s_1) P_a^2(\s_2|\s_1) \cdots
    P_a^N(\s_N | \s_{N-1},\cdots, \s_1) \ .
\end{equation}
Each term $P_a^i$ is then approximated by a neural network, which takes as input $\{\s_1,\cdots,\s_{i-1}\}$ and gives as output $P_a^i$, i.e. the probability of $\s_i$ conditioned to the input. 
Such a representation of $P_a$, also called Masked Autoencoder for Distribution Estimator (MADE)~\cite{Germain}, 
allows for very efficient sampling, because one can first sample $\s_1$, then $\s_2$ given $\s_1$, and so on, in a time scaling as the sum of the computational complexity of evaluating each of the $P_a^i$, which is typically polynomial in $N$ for reasonable architectures. Hence, this scheme satisfies condition (i) above.
The simplest choice for such a neural network is a linear layer followed by a softmax activation function. 
Ref.~\cite{Wu2019} showed that using such an architecture, several statistical models could be well approximated, and the Boltzmann distribution of a Sherrington-Kirkpatrick (SK) spin glass model (with $N=20$) could be efficiently sampled. Note that the model in Ref.~\cite{Wu2019} was trained by a {\it variational} procedure, which minimizes
$D_{KL}(P_a||P_t)$ instead of $D_{KL}(P_t||P_a)$. This method is computationally very efficient as it only requires an average over $P_a$, which can be sampled easily, instead of $P_t$, but it is prone to mode-collapse (see Sec.~\ref{sec:train} for details). Moreover, this work was limited to quite small $N$.

Following up on Ref.~\cite{Wu2019}, Ref.~\cite{PhysRevE.101.053312} considered as target probability the Boltzmann distribution of a two-dimensional (2d) Edwards-Anderson (EA) spin glass model at various temperatures $T$, and used a Neural Autoregressive Distribution Estimator (NADE)~\cite{Uria2016}, which is a variation of the MADE meant to reduce the number of parameters. Furthermore, the model was trained using a different scheme from Ref.~\cite{Wu2019}, called {\it sequential tempering}, which tries to minimize $D_{KL}(P_t||P_a)$, thus preventing mode-collapse. To this aim, at first, a sample from $P_t$ is generated at high temperature, which is easy, and used to learn $P_a$. Then, temperature is slightly reduced and smart MCMC sampling is performed using the $P_a$ learned at the previous step, to generate a new sample from $P_t$, which is then used in the next step. If $P_a$ remains a good approximation to $P_t$ and MCMC sampling is efficient, this strategy ensures a correct minimization of $D_{KL}(P_t||P_a)$. This was shown to be the case in Ref.~\cite{PhysRevE.101.053312}, down to low temperatures for a 2d EA model of up to $N=225$ spins. 

Ref.~\cite{gabrie2021adaptive} introduced a different scheme for learning $P_a$. This {\it adaptive} scheme combines local MCMC moves with smart $P_a$-assisted MCMC moves, together with an online training of $P_a$. It was successfully tested using a different architecture for $P_a$ (called normalizing flows), on problems with two stable states separated by a high free energy barrier. Note that normalized flows can be equivalently interpreted as autoregressive models~\cite{NIPS2017_6c1da886,NIPS2016_ddeebdee,Hartnett20}. 
Ref.~\cite{wu21} also proved the effectiveness of smart assisted MCMC moves in a 2d Ising model and an Ising-like frustrated plaquette model.

\begin{figure*}[t]
    \centering
    \includegraphics[width=\textwidth]{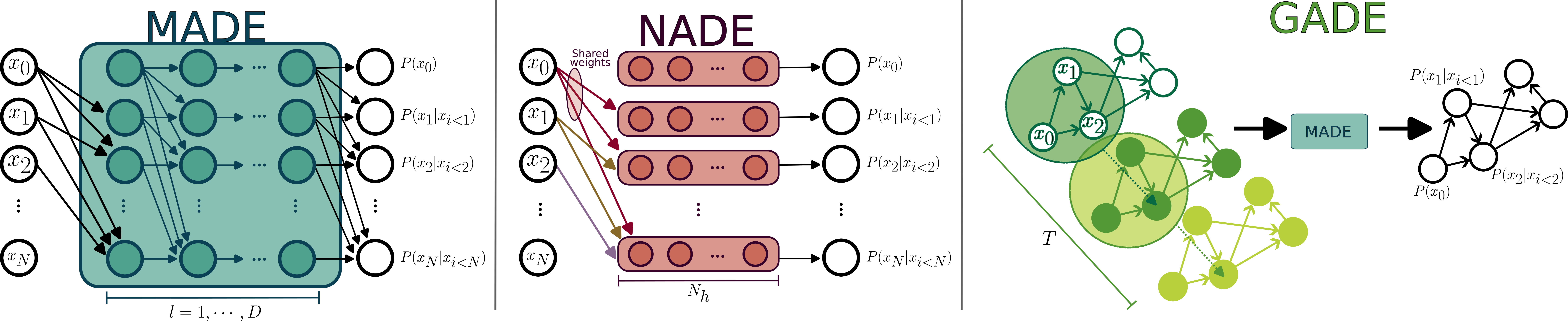}
    \caption{Sketches of the autoregressive architectures used in this work.}
    \label{fig:architectures}
\end{figure*}

Several other groups~\cite{Hibat-Allah2021,fan2021finding,schuetz2022graph,condmat7020038} investigated a problem related to sampling, namely that of simulated annealing~\cite{KGV83} for finding ground states of optimization problems. 
This is an {\it a priori} slightly easier problem, because simulated annealing does not need to equilibrate at all temperatures to find a solution~\cite{KK07,KZ13}.
In these works, simulated annealing moves were once again assisted by machine learning. Ref.~\cite{Hibat-Allah2021} tested their procedure on the 2d EA and SK models, and Ref.~\cite{fan2021finding} considered a 2d, 3d, and 4d EA model. However, while finding the exact ground state of the SK and EA (for $d\geq 3$) models is hard, in practice for not too large random instances the problem can be solved by a proper implementation of standard simulated annealing~\cite{Wang2015}, and the scaling of these methods with system size remains poorly investigated.
Ref.~\cite{schuetz2022graph} considered the graph coloring problem, which is the zero-temperature version of the benchmark problem we propose to use in this work, and found that a Graph Neural Network (GNN) can propose moves that allow one to efficiently find a proper coloring with comparable performances to (but not outperforming) state-of-the-art local search algorithm.
Additionally, GNN have shown to be successful at solving discrete combinatorial problems~\cite{schuetz22}, but they do not provide much advantage over classical greedy algorithms, and sometimes they can even show  worse performance~\cite{Angelini22,boettcher22}.
Finally, Ref.~\cite{condmat7020038} showed that the machine-learning-assisted simulated annealing scheme does not work on a glassy problem with a rough energy landscape.

These works provided a series of inspiring ideas to improve sampling in disordered systems via machine learning smart MCMC moves. Yet, the question of whether machine learning can really speed up sampling in problems that are exponentially hard to sample via local MCMC remains open. This wide class of systems include many problems of interest, such as optimization problems (e.g. random SAT or random graph coloring)~\cite{KMRSZ07} and mean-field glass-forming materials~\cite{KW87,KT88,CK93,parisi2020theory}.

\subsection{Summary}

In this work, we test machine-learning-assisted MCMC in what is considered to be a prototypical hard-to-sample model, namely the coloring of random graphs~\cite{cheeseman1991really,mulet2002coloring,Zdeborova2007}. Before doing that, we also tested and reproduced previous results in simpler cases. 

The models we consider are:
\begin{itemize}
    \item[(1)] The mean-field ferromagnetic problem, usually called the Curie-Weiss (CW) model, to gain some analytical insight into the different ways of training the auxiliary model.
    \item[(2)] A two-dimensional Edwards-Anderson spin glass (2d EA) model. We consider this as an `easy' problem (because, for instance, its ground state can be found in polynomial time), and we use it to reproduce previous results, compare different architectures, and gain insight on the role of some hyperparameters. \item[(3)] The coloring (COL) of a random graph, which at finite temperature becomes an antiferromagnetic Potts model. In the proper range of parameters, this problem is proven to be exponentially hard to sample via local MCMC~\cite{MS06,Zdeborova2007}, and we use it as a benchmark to understand whether smart MCMC can improve the sampling efficiency.
\end{itemize}
Any machine learning model that satisfies the autoregressive property can be trained and used as an auxiliary distribution to propose smart moves. However, on the one hand, for complex problems, shallow or simple models might not be expressive enough to accurately learn the target distribution. On the other hand, if a problem can be easily solved by a simpler model, there is no need to employ complex deep architectures. 
In this paper we used several standard architectures illustrated in Fig.~\ref{fig:architectures} and detailed in the SI: 
\begin{itemize}
    \item The MADE~\cite{Wu2019}, which is an autoregressive deep neural network;
    when its depth is equal to zero, this corresponds to a `shallow' or single-layer autoregressive model.
    \item The NADE~\cite{Uria2016,PhysRevE.101.053312}, which corresponds to a MADE with additional constraints on the parameters, with different depths and number of hidden units.
    This architecture has proven to be effective in image detection~\cite{Hashemzehi2020}, filtering~\cite{pmlr-v48-zheng16} and quantum systems~\cite{Sharir2020}.
    \item For the coloring, because neither the MADE nor the NADE perform well, we also tested  an autoregressive GNN that we called Graph Autoregressive Distribution Estimator (GADE), and a non-symmetric MADE (called ColoredMADE). 
\end{itemize}
Finally, we tried several strategies to learn the auxiliary model:
\begin{itemize}
    \item[(I)] {\it Maximum likelihood}: we generate a sample from the target distribution, and we use it to train the auxiliary model by maximum likelihood. While this is not a technique to generate samples from $P_t$, because one needs the samples to begin with, it is the best way to test if a given architecture for $P_a$ is expressive enough.
    \item[(II)] {\it Variational}: we minimize the KL divergence $D_{KL}(P_a||P_t)$, which also corresponds to the variational free energy of the auxiliary model when considered as an approximation of the true one~\cite{Wu2019}.
    \item[(III)] {\it Sequential tempering}: we train the auxiliary model at higher temperature $T$, then use it to generate samples at lower $T$, and use the new samples to re-train the auxiliary model, and so on~\cite{PhysRevE.101.053312,gabrie2021adaptive}. 
\end{itemize}
The core of our work is the application of methods (II) and (III) to attempt a sampling at low temperatures, for which local MCMC does not decorrelate fast enough.
For the 2d EA model, we find that basically all the techniques and architectures perform well down to very low temperatures, although (II) is more prone to mode-collapse. We confirm that machine learning MCMC moves can provide a speedup in this case~\cite{PhysRevE.101.053312}.
     For the COL problem, we find that none of these methods work well, even at moderately high temperatures located within the paramagnetic phase of the model.

\section{Methods}
\label{sec:train}

We consider a specific application of the general scheme discussed in Sec.~\ref{sec:intro}, in which
we want to approximate the Boltzmann distribution associated to the `true' Hamiltonian $H(\s)$ at a fixed inverse temperature $\beta=1/T$
(the target),
\begin{equation}
    P_t(\s) = P_B(\s) = \frac{e^{-\beta H(\s)}}{Z} \ ,
\end{equation}
with an autoregressive (AR) network, i.e. ${P_a(\s)=P_{AR}(\s)}$. In most cases, sampling is easy at small $\beta$ and becomes harder and harder as $\beta$ is increased.
We now discuss different possible strategies to learn a proper $P_{AR}(\s)$.

\subsection{Maximum likelihood}
\label{sec:maxlike}

If a sufficiently large sample of configurations $\{\s^m\}_{m=1,\cdots,M}$ is available, independently and identically sampled from $P_B(\s)$, it is possible to train the model by maximizing the probability of the sample according to the AR model itself, i.e.~to use the maximum likelihood method.

Assuming that the AR model is specified by a set of parameters $\theta$, we maximize the likelihood of the observed data, defined as
\begin{equation}
    \mathcal{L}(\theta) = \prod_{m=1}^M P_{AR}(\s^m|\theta) \ .
\end{equation}
Equivalently, if $P_{emp}(\s)=\frac1M\sum_{m=1}^M \delta_{\s,\s^m}$ is the empirical distribution of the sample, we minimize
\begin{equation}\label{eq:DKLemp}
\begin{split}
    {\cal D}(\theta) &= D_{KL}(P_{emp}||P_{AR}) \\ &= 
    -\log M -\frac{1}{M} \sum_{m=1}^M \log P_{AR}(\s^m|\theta) \ .
\end{split}\end{equation}
The optimal parameters $\hat{\theta}$ are given by 
\begin{equation}
\hat{\theta} = \mathrm{argmax}_{\theta} \mathcal{L}(\theta)
= \mathrm{argmin}_{\theta} \mathcal{D}(\theta) \ .
\end{equation}
The gradient of $\mathcal{D}(\theta)$ can be computed analytically in terms of the $\s^m$, because $\log P_{AR}(\s|\theta)$ is given explicitly (or by back-propagation) as a function of $\theta$ at fixed $\s$, and the training can thus be performed efficiently.

Note that if $P_B(\s)$ has multiple peaks, for sufficiently large $M$ all these peaks are represented in the empirical distribution with the correct weights. Hence, when performing maximum likelihood to learn $P_{AR}(\s)$, the learned model should be able to represent all the peaks of $P_B(\s)$, provided the AR model has enough free parameters, i.e. is expressive enough.
Then, mode-collapse will not occur.

Obviously, the maximum-likelihood approach relies on the quality of the initial sample, which has to be representative of the true distribution. Such a sample is by definition difficult to obtain for the really hard sampling problems that we want to solve. Moreover, if we were able to obtain such a sample by conventional means, there would be no need for any smart MCMC scheme. Nevertheless, the maximum-likelihood approach constitutes a reliable and effective way to test if a specific AR architecture is capable of learning the complexity of a specific problem. We will thus use this scheme, in cases where standard sampling from $P_B(\s)$ is possible, to test the expressive quality of our AR architectures.

\subsection{Variational approach}
\label{sec:var}

Ref.~\cite{Wu2019} proposes to 
bypass the need for sampling from $P_B(\s)$ 
by using a variational approach.
Instead of minimizing
$D_{KL}(P_{B}||P_{AR})$ or its approximation $D_{KL}(P_{emp}||P_{AR})$, as in Sec.~\ref{sec:maxlike},
we want here to minimize $D_{KL}(P_{AR}||P_B) = \sum_\s P_{AR}(\s) \log \frac{P_{AR}(\s)}{P_B(\s)}$, or equivalently~\cite{Wu2019} the variational free energy:
\begin{equation}\begin{split}
    \beta F[P_{AR}] &= \sum_\s P_{AR}(\s)(\beta H(\s) +\log P_{AR}(\s)) \\
    & = \beta F[P_B] + D_{KL}(P_{AR}||P_B)
    \ .
\end{split}\end{equation}
As it is well known in statistical mechanics, because the KL divergence is positive, $F[P_{AR}]$ is minimized when $P_{AR}=P_B$ and $D_{KL}(P_{AR}||P_B)=0$, and otherwise it provides an upper bound to the true free energy.

The gradient with respect to the parameters $\theta$ that define the AR model can be written as an expectation value over the AR model itself~\cite{Wu2019},
\begin{equation}\label{eq:gradvar}
\begin{split}
    \beta \nabla_\theta F[P_{AR}] &= \langle  Q(\s) \nabla_\theta \log P_{AR}(\s) \rangle_{P_{AR}} \ , \\
    Q(\s) &= \beta H(\s) + \log P_{AR}(\s) \ .
\end{split}\end{equation}
The learning can then be done by gradient descent on $F[P_{AR}]$,  sampling from the AR model to estimate the gradient via Eq.~\eqref{eq:gradvar}.
We used, as a condition to stop the gradient descent, that the variance of $Q(\s)$ over batches of generated data is smaller than a given threshold. Indeed, if the AR distribution is exactly the Boltzmann one, then $Q(\s)$ is a constant. Reciprocally, if the variance of $Q(\s)$ is zero, then the AR distribution is proportional to the Boltzmann distribution whenever $P_{AR}(\s)>0$, but not necessarily over all possible $\s$, due to mode-collapse. 
More specifically, if we have mode-collapse, then $P_{AR}(\s) = P_B(\s)/\mathcal{Z}$ in some regions of the space of $\s$ (typically around some of the peaks of $P_B$) and $P_{AR}(\s)=0$ elsewhere, where the proportionality constant $\mathcal{Z}<1$ represents the total probability covered by $P_{AR}(\s)$ in $P_B(\s)$.
We then obtain $D_{KL}(P_{AR}||P_B)=-\log\mathcal{Z}>0$, because the regions where $P_{AR}(\s)=0$ do not contribute to the sum.
While this solution has larger KL divergence with respect to the optimal one ($P_{AR}=P_B$, $D_{KL}(P_{AR}||P_B)=0$), it can be a local minimum of $D_{KL}(P_{AR}||P_B)$ in which the variationally-trained AR model can be trapped, thus learning only a part of the landscape.

\subsection{Local versus global MCMC}
\label{sec:loc_vs_glo_MCMC}

Standard local MCMC usually consists in selecting a variable at random and then proposing a random (or semi-random) change.
If the MC moves respect microscopic detailed balance, the MCMC is guaranteed to converge to the correct Boltzmann distribution.
This can be achieved, e.g., by accepting/rejecting the proposed MC moves following the Metropolis rule, where the acceptance probability of a move from configuration $\s_{old}$ to $\s_{new}$ is defined as:  
\begin{equation}
    \mathrm{Acc}\left[ \s_{old}\xrightarrow{} \s_{new}\right] = \min \left[ 1,\frac{P_{B}(\s_{new})}{P_{B}(\s_{old})} \right] \ .
\end{equation}
We call this scheme local MCMC because each move consists in a change
of a single degree of freedom.

In contrast, we can sample from our autoregressive model $P_{AR}(\s)$ to generate a new proposed configuration $\s_{new}$. 
It is still useful to respect microscopic detailed balance in order to ensure convergence to equilibrium, and for this reason the replacement $\s_{old} \to \s_{new}$ is accepted with probability
\begin{equation}\label{eq:ARglobal}
    \mathrm{Acc} \left[ \s_{old}\xrightarrow{} \s_{new}\right]= \min \left[ 1,\frac{P_{B}(\s_{new}) \times P_{AR}(\s_{old})}{P_{B}(\s_{old})\times P_{AR}(\s_{new})} \right] \ .
\end{equation}
Note that, because $\s_{new}$ is generated from scratch by the AR model, it is in most cases completely different from $\s_{old}$, hence the resulting move is non-local and this is why this scheme is called global MCMC. 

We also note that this global MCMC scheme is very similar to importance sampling, in which $M$ i.i.d. samples $\s^m$ are generated from $P_{AR}(\s)$, and then reweighted by $W(\s^m) = P_{B}(\s^m)/P_{AR}(\s^m)$ to compute averages. However, the formulation in terms of a MCMC is convenient to compare with local MCMC, to monitor efficiency via the acceptance rate (which is morally equivalent to a participation ratio in importance sampling), and to perform smart protocols (e.g. sequential tempering) during the MCMC dynamics~\cite{PhysRevE.101.053312,gabrie2021adaptive}. This is why we stick to this formulation in this paper.

The reweighting factor $W(\s) = P_{B}(\s)/P_{AR}(\s)$ that appears both in importance sampling and in Eq.~\eqref{eq:ARglobal} is the crucial quantity for the efficiency of the global MCMC scheme. If $W(\s)$ is typically exponential in $N$, then its fluctuations are too wild and moves are almost never accepted. The KL divergence is precisely the average of $\log W(\s)$, either on the Boltzmann or on the AR distribution, and if it is too large (in particular, growing linearly in $N$) the method is doomed to failure.

\subsection{Sequential tempering}
\label{sequential_temp}

Sequential tempering is
a technique used in Ref.~\cite{PhysRevE.101.053312}
to learn the AR probability at a larger $\beta$ using data from lower $\beta$, which can be convenient because collecting data becomes harder upon increasing $\beta$.
The first step consists in collecting a sample via local MCMC at low $\beta$, where sampling is easy, and then training an AR model to reproduce this sample by maximum likelihood (Sec.~\ref{sec:maxlike}). Next, in order to create a new sample at $\beta +\delta \beta$, we use global MCMC by proposing moves with the previous AR model, at the new temperature. The acceptance rule then becomes:
\begin{equation}
    \mathrm{Acc} \left[ \s_{old}\xrightarrow{} \s_{new}\right]= \min \left[ 1,\frac{e^{-(\beta + \delta \beta) H(\s_{new})} P_{AR}(\s_{old})}{e^{-(\beta + \delta \beta) H(\s_{old})} P_{AR}(\s_{new})} \right] \ .
\end{equation}
We then learn a new AR model from the new sample by maximum likelihood, and iterate until either we reach the temperature of interest, or the convergence time of the global MCMC exceeds some fixed threshold, indicating a failure of the training procedure.
A related adaptive global MCMC scheme has been introduced in Ref.~\cite{gabrie2021adaptive} and is detailed in the SI.

\subsection{Evaluation of the AR model}
\label{sec:evaluation}

Once the learning is completed, one can use several observables to evaluate the quality of the AR model. By completion of the learning we mean convergence of the gradient ascent in maximum likelihood (Sec.~\ref{sec:maxlike}), convergence of the gradient descent in the variational free energy (Sec.~\ref{sec:var}), or reaching the target temperature with high enough acceptance rate in the sequential tempering (Sec.~\ref{sequential_temp}).

As a first check, we can use the AR model to estimate thermodynamic observables (energy, entropy, correlations) of the true Hamiltonian.
If the AR model has a lower entropy than the true one, the AR model is probably suffering mode-collapse. 
Another interesting observable is the KL divergence, either $D_{KL}(P_{AR}||P_B)$ or $D_{KL}(P_{B}||P_{AR})$, which measures how well the AR model approximates the target one, and more quantitatively provides the average of the reweighting factor, as discussed in Sec.~\ref{sec:loc_vs_glo_MCMC}. 
A more easily accessible quality measure consists in generating samples with the AR model, then evolving them with local MCMC and checking if the energy remains constant and the correlation functions remain time-translationally invariant~\cite{CK93,NBC17}, as expected if the initial configuration is a good equilibrium one.

A very important measure of the quality of the AR model is the acceptance rate of the global MCMC, as a high acceptance rate indicates that the AR model describes well the true model, at least in the region explored by the global MCMC. Yet,
the AR model could be mode-collapsed and still keep a high acceptance rate, because the global MCMC would then only explore the region on which the AR model has collapsed~\cite{gabrie2021adaptive}. In particular, this could happen if:
\begin{itemize}
    \item The true model has a few (or a polynomial number of) probability peaks. In that case, the AR model could have learned the correct energy and correct free energy, and still the entropy could be lower than the true one, but by a sub-extensive (hence difficult to detect) amount. This is the standard scenario of mode-collapse, e.g. in a ferromagnetic model with two states, when the AR model is only learning one of them.
    \item The true model has a number of peaks growing exponentially in $N$, hence associated to an extensive contribution to the entropy (usually called configurational entropy or complexity in the glass literature~\cite{KMRSZ07}). In this case, if the AR model has learned the true energy (or a higher one), and its entropy is much lower than the true one, then its free energy should be appreciably higher than the true one. 
    If instead the AR model with mode-collapse learns a good approximation to the true free energy, this happens by lowering both the entropy and the energy. In this case, mode-collapse is easily detected by a drift of the energy, that increases with the number of iterations when MCMC is started from AR-generated samples. 
\end{itemize}

In practice, if (i)~the global MCMC has good acceptance rate, (ii)~the global MCMC dynamics initialized in a configuration generated by the AR model is close to stationary (i.e. one-time quantities such as the energy are constant in time, and two-time quantities such as correlations only depend on the time difference), and (iii)~the time-dependent correlation of global MCMC vanish for large time differences,
then the model should be of high enough quality. 
Yet, ultimately, it is hard to assess the quality of the AR model in a general way. We will give more concrete examples in the rest of the paper.


\section{Curie-Weiss model}

As a first exercise, we discuss here a mean-field ferromagnetic model, namely the Curie-Weiss (CW) model, which provides some insight on the expressive properties of AR models.

\begin{figure*}[t]
\includegraphics[width=0.8\textwidth]{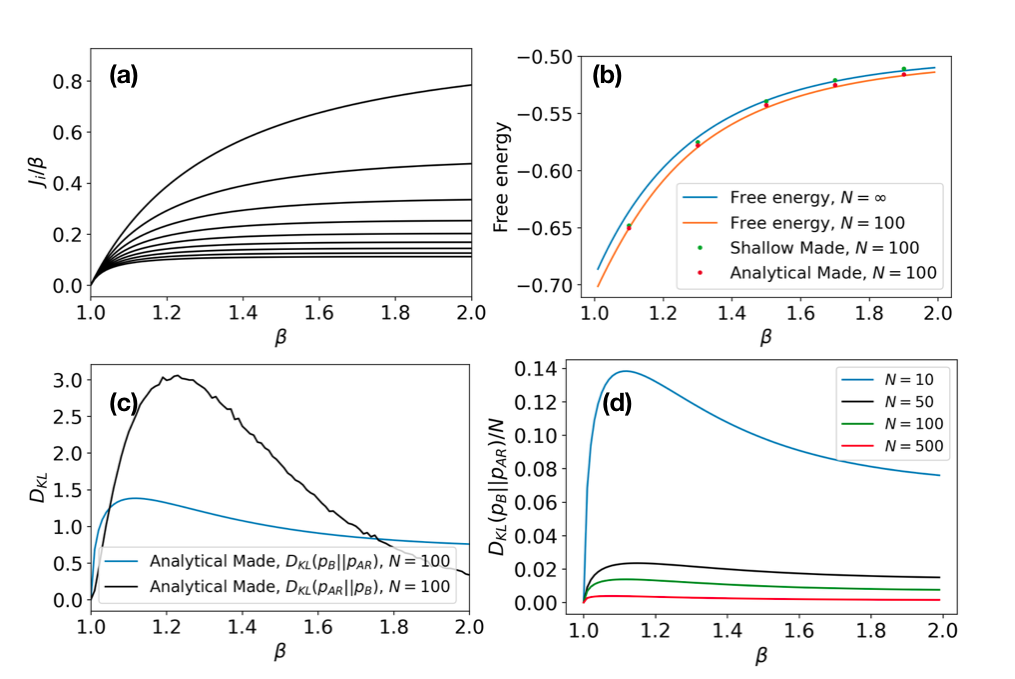} 
  \caption{{\bf Curie-Weiss model -} (a)~Analytical solution of Eq.~\eqref{eq:CWJi} for the first nine couplings $J_2, J_3, \cdots, J_{10}$ (from top to bottom, rescaled by inverse temperature $\beta$) of the autoregressive model as a function of $\beta$, in the thermodynamic limit $N\to\infty$.
(b)~Comparison of the exact free energy per spin, obtained from the exact solution for $N\to\infty$ (full blue line) or for $N=100$ (full orange line), 
  with its estimate obtained from two autoregressive models at $N=100$, either a shallow MADE trained via maximum likelihood on equilibrium samples generated via local MCMC on the true Boltzmann distribution of the CW model (green circles), or the shallow MADE obtained from the solution of Eq.~\eqref{eq:CWJi} (orange circles).
  (c)~KL divergences between the shallow MADE obtained from Eq.~\eqref{eq:CWJi} and the target model, computed for $N=100$ as a function of $\beta$.
(d)~Scaling of the KL divergence with~$N$.
  }
  \label{analytic}
  \label{analytic2}
\end{figure*}

\subsection{The model and its Boltzmann distribution}

The model is defined on a set of binary Ising spins, for convenience $\s_i \in \{ -1, 1\}$, by the Hamiltonian
\begin{equation}
    H(\s) = - \frac{N}{2} m^2 \ , \quad m = \frac{1}N \sum_i \s_i \ .
\end{equation}
In the thermodynamic limit, the value of $m$ that dominates the Boltzmann probability distribution
is the solution of $m = \tanh(\beta m)$ (see the SI for details),
which gives $m=0$, i.e. a paramagnetic phase, for $T>1$, and $m =\pm \hat m$, i.e. a ferromagnetic phase, for $T<1$. Note that the two solutions with positive or negative $m$ have the same free energy, hence the probability distribution of the model has two distinct peaks corresponding to these two ferromagnetic states.
The model has a spin-flip symmetry $\s_i\to -\s_i$, which is spontaneously broken for $T<1$.

For $N\to \infty$ and $T>1$, the Boltzmann probability distribution is well approximated by a product distribution,
\begin{equation}\label{eq:CWpara}
P_B(\s)=\frac{e^{-\beta H(\s)}}Z \sim \prod_i p(\s_i) \ , \quad p(\s_i) = \frac{1}{2} \ ,
\end{equation}
in the sense that the KL divergence (per spin) between the two sides of Eq.~\eqref{eq:CWpara} vanishes for $N\to\infty$.
When $T<1$, this holds separately for each of the two ferromagnetic states, resulting in a mixture of two product distributions:
\begin{equation}\label{eq:CWfact}
\begin{split}
P_B(\s) &\sim \frac{1}2 \prod_i p_+(\s_i) + \frac{1}2 \prod_i p_-(\s_i) \ , \\
p_\pm(\s_i) &= \frac{1 \pm \hat m \s_i}{2} \ ,
\end{split}\end{equation}
each peak being approximated by the product form with $m=\hat m$ and $m=-\hat m$. 

\subsection{Expressivity of the autoregressive model}

The CW model is well approximated by a mixture of two product distributions for large $N$. We now want to ask whether this structure can also be well approximated by a single AR model.
Because training an AR model on a sample generated by the CW model is only possible numerically, we consider instead a sample generated by the mixture model, which coincides with the CW one for $N\to\infty$. For this model, we can provide an analytic solution.

We then assume that an infinite sample of spin configurations is generated from the mixture model in Eq.~\eqref{eq:CWfact}, i.e. with probability $1/2$ we choose $m=\pm \hat m$ and we then choose each spin independently with probability $p_{\pm}(\s_i)$.
We use this infinite sample to train the AR model via maximum likelihood (Sec.~\ref{sec:maxlike}).
We consider here a shallow AR model (see the SI for details).
Because of the spin-flip symmetry, which is realized in an infinite sample from the true model, the local fields $h_i=0$ have to vanish. Furthermore,
because the Hamiltonian is invariant under any relabeling of the $\{\s_i\}$, in each conditional probability $P^i_{AR}(\s_i|\s_{i-1},...,\s_1)$ the spins $\s_{i-1},...,\s_1$ are seen as equivalent by spin $\s_i$,
hence the couplings $J_{ij}$ should be independent of the index $j<i$, leading to
\begin{equation}\label{eq:CW-ARmaxlike}
P^i_{AR}(\s_i|\s_{<i}) = \frac{\exp(\sum_{j(<i)} J_i \s_i\s_j)}{2\cosh ( \sum_{j(<i)} J_i \s_j )}
=\frac{\exp(J_i \s_i m_{<i})}{2\cosh ( J_i m_{<i} )}
\ ,
\end{equation}
where $m_{<i} = \sum_{j(<i)} \s_j$ is the magnetization of the first $(i-1)$ spins.
Hence, each term $P^i_{AR}$ is parametrized by a single coupling $J_i$.

Minimization with respect to $J_i$
of the KL divergence between the mixture model 
in Eq.~\eqref{eq:CWfact} and the AR model in Eq.~\eqref{eq:CW-ARmaxlike}
leads to a set of coupled equations:
\begin{equation}\label{eq:CWJi}
    \begin{split}
    p(k)&={i-1\choose k}\left(\frac{1+\hat m}{2}\right)^k \left(\frac{1-\hat m}{2}\right)^{i-1-k} \ ,  \\
        \hat m^2 &= \sum_{k = 0}^{i-1} p(k) \frac{2k-i+1}{i-1}\tanh[J_i(2k-i+1)] \ .
    \end{split}
\end{equation}
Once $\hat m$ is computed by solving $m=\tanh(\beta m)$, the $J_i$s can be computed by solving Eq.~\eqref{eq:CWJi} numerically. 
We can also compute the resulting minimum KL divergence (SI).

Because $\hat m=0$ for $T>1$, all the $J_i=0$ and 
the AR model trivially reduces to the product distribution in Eq.~\eqref{eq:CWpara}.
Results for $T<1$ are shown in Fig.~\ref{analytic}a for the first few couplings.
While we find a non-trivial relation between $J_i$ and $\beta$ in the vicinity of the critical temperature $\beta=1$ where all the $J_i$ vanish, Fig.~\ref{analytic}a suggests that $J_i$ roughly scales as $\beta/(i-1)$ at large $\beta$.
A comparison of the free energy per spin, obtained from the exact solution and from the AR model, is given in Fig.~\ref{analytic2}b.
The KL divergences between $P_{AR}$ and $P_{B}$ (in both directions) are given in Fig.~\ref{analytic2}c; note that these are not divided by $N$, hence their value per spin is very small, and shows a mild peak in the vicinity of the phase transition.
Finally, Fig.~\ref{analytic2}d shows that $D_{KL}(P_{B}||P_{AR})/N$ vanishes at all temperatures when $N\to\infty$, which confirms that the AR model can perfectly approximate the target one in the thermodynamic limit, even in the ferromagnetic phase.

This calculation gives some useful insight on the expressivity and training of AR models:
\begin{itemize}
    \item If training is done via maximum likelihood with a perfect sample of the target model, no mode-collapse is observed, as expected.
\item A single shallow AR model is capable of expressing a two-peaked distribution, at the price of having finite couplings that are much stronger than the CW ones (which where of order $\beta/N$, taking into account the factor of $\beta$ in the Boltzmann measure). This happens as follows. The first spin $\s_1$ is chosen at random, $P^1_{AR}(\s_1)=1/2$. The second spin has a strong coupling with the first, $J_2 \sim \beta$. The third spin also has a strong coupling with the first two, $J_3\sim \beta/2$, and so on. More generally, this suggests that for $q$-state variables, a single AR model can express a distribution with $q$ peaks, each peak being selected by the choice of the first spin. Whether an AR model can express a distribution with more than $q$ peaks remains open at this stage, but we will see in Sec.~\ref{sec:col_res} that this is not possible when there are too many peaks. 
\item The maximum-likelihood training converges to a solution at all temperatures, and does not suffer from any slowdown in the vicinity of the critical point. 
\end{itemize}
We thus conclude that AR models can represent well models with second-order phase transitions, without suffering from slowing down, at least when training is done via maximum likelihood.

\subsection{Mode-collapse}
\label{sec:CWmodecollapse}

We also consider a shallow AR model trained in a variational way (Sec.~\ref{sec:var}) to represent the CW model.
In this case, we do not impose any restriction on the couplings $J_{ij}$ and the fields $h_i$ (see the SI), and the variational free energy cannot be written explicitly.
For $T<1$, we found numerically that the AR model can end up in three different fixed points. The first one corresponds to the `correct' form in Eq.~\eqref{eq:CW-ARmaxlike}, i.e. with $h_i=0$ and $J_{ij}=J_i$. The other two correspond to mode-collapse in a solution with $J_{ij}=0$ and $h_i=h= \pm T \text{atanh}(\hat m)$: in each of these two solutions, the AR model learns only one of the two peaks of the Boltzmann distribution, resulting in a loss of entropy per spin of $(\log 2)/N$, which is however too small to be detected for large $N$.
Moreover, the basin of attraction of the correct solution seems to be vanishingly small for large $N$, and the two mode-collapsed solutions dominate the training. In fact, to obtain the solution with $h_i=0$, we have to force the fields to vanish, thus imposing the spin-flip symmetry.
We conclude that, as expected, one has to be careful about mode-collapse when training the AR model variationally.


\section{Edwards-Anderson model}
\label{sec:EA}

We now consider a two-dimensional (2d) Edwards-Anderson (EA) spin glass model, chosen because it is not exponentially hard to sample (finding its ground state requires polynomial time in $N$~\cite{barahona1982computational,hartmann2011ground,CJMM22}), 
there is no phase transition at finite temperature (hence no {\it a priori} risk of mode-collapse),
and yet the relaxation time of local MCMC grows quite fast upon lowering temperature. Moreover, previous work~\cite{PhysRevE.101.053312} has explored this model, finding an important speedup using a NADE architecture.

\subsection{Model and architectures}
\label{sec:2dEAmethods}

We consider a 2d EA model defined on a $10\times 10$ square lattice with periodic boundary conditions, i.e. with ${N=100}$ Ising spins, $\s_i \in \{-1,1\}$, as
\begin{equation}
    H(\s) = -\sum_{\langle ij \rangle} J_{ij}\s_i\s_j \ .
\end{equation}
Here, the sum runs over neighboring pairs of sites in the square lattice, and the
couplings $J_{ij}$ are i.i.d. from a Gaussian distribution with zero mean and unit variance as in Ref.~\cite{PhysRevE.101.053312}.

\begin{figure*}[t]
  \centering
  \includegraphics[width=1\textwidth]{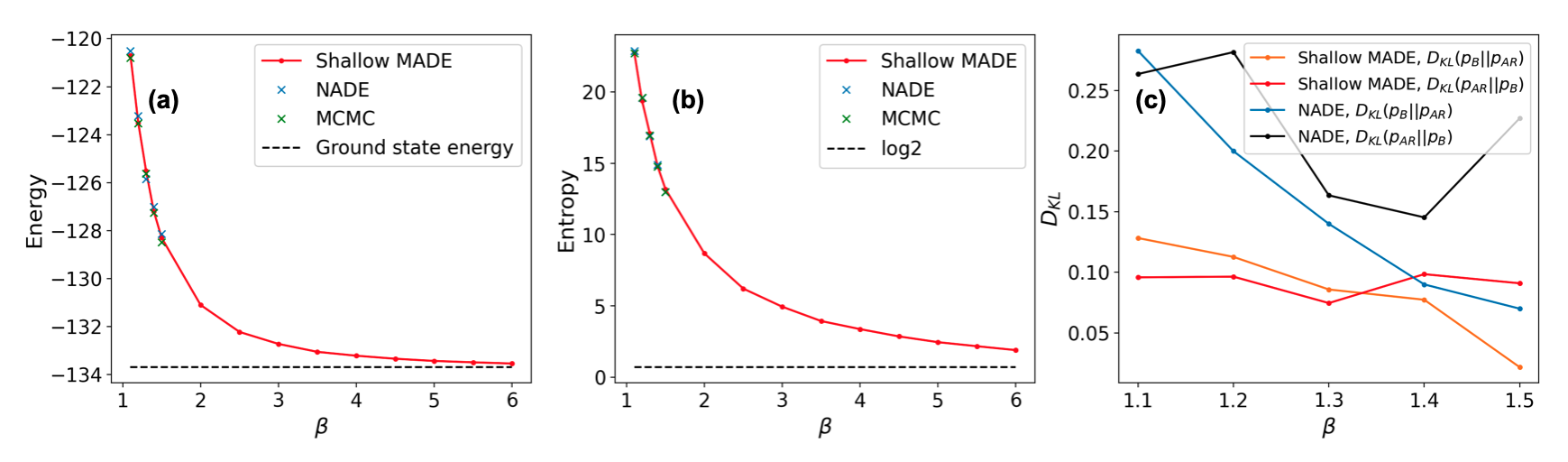}  
  \includegraphics[width=1\textwidth]{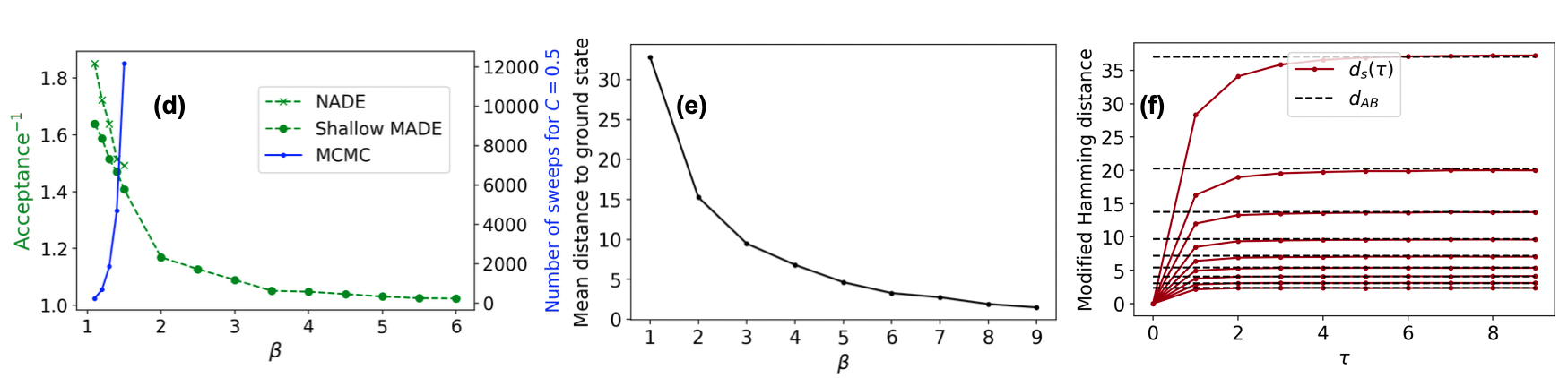}  
    \caption{{\bf 2d Edwards-Anderson model} with $N=100$ spins ($10\times 10$ periodic square lattice) and a training sample of $M=10^5$ configurations. (a)~Total energy and (b)~total entropy as a function of $\beta$, as estimated by  the $N_h = 64$ NADE and by the shallow MADE, compared with the exact result obtained by local MCMC. The entropy is obtained by thermodynamic integration, and the exact ground state energy has been obtained via the McGroundstate server~\cite{CJMM22}.
    (c)~KL divergences between the AR models and the Boltzmann distribution as a function of $\beta$.
(d)~Inverse of the acceptance rate of global MCMC as a function of inverse temperature $\beta$, for the $N_h = 64$ NADE and the shallow MADE, both trained via sequential tempering.  For comparison, we also show the decorrelation time of local MCMC.
(e)~Average Hamming distance (i.e. number of flipped spins) between the ground state and equilibrium configurations generated by global MCMC with the shallow MADE, as a function of $\beta$.
(f)~Average Hamming distance $d_{AB}$ between two independent global MCMC runs with the shallow MADE (dashed horizontal lines), compared with the average Hamming distance $d_s(\tau)$ between a single global MC chain at two distinct times separated by $\tau$ global MC steps (red dots and lines). From top to bottom, $\beta=1,2,3,\cdots,9$.
}  
\label{ener_entro_EA}
\end{figure*}

For the AR decomposition in Eq.~\eqref{eq:BayesAR},  
we order the variables moving along e.g. horizontal lines from the top left to the bottom right of the lattice.
We consider two different AR architectures:
a shallow MADE without fields ($h_i=0$) and  with $N(N-1)/2$ independent coupling 
parameters $J_{ij}$, 
and a NADE model with $N_h$ hidden nodes (see the SI). In the NADE, in order to impose the inversion symmetry $\s_i\to-\s_i$, we set all the biases to zero, and we use sigmoid functions in the hidden layers.
The NADE then has $2\times N_h \times N$ parameters. 

\begin{table}[b]
\begin{tabular}{ |c|c|} 
 \hline
 Local MCMC time for one sweep & $10^{-5}$ s \\ 
 Global MCMC time for one move & $3 \cdot 10^{-4}$ s  \\ 
 Total learning time at given $\beta$ & $5$ mn \\ 
 \hline
\end{tabular}
\caption{Computational time needed to perform local MCMC, global MCMC, and model learning, on a standard  laptop.}
\label{tab:EA}
\end{table}

In order to train the two AR models, we use sequential tempering (Sec.~\ref{sequential_temp}) as in Ref.~\cite{PhysRevE.101.053312}.
The initial sample used to learn the model by maximum likelihood is generated at $\beta = 1$, where local MCMC still decorrelates fast enough. It contains $M=10^5$ configurations, and is obtained by a single local MCMC, adding a new configuration every $10$ sweeps, resulting in a total of $10^6$ sweeps. The models are learned with gradient descent using mini-batches of $256$ configurations, during $50$ training epochs and with a learning rate of $10^{-3}$. 
During the sequential tempering, each sequence of the sample is evolved for $10$ global MCMC steps, at a new inverse temperature $\beta \leftarrow \beta +0.1$. The acceptance rate is also computed during this global MCMC. The model is then trained again, with the same hyperparameters, using the new sample.
The times needed on a standard laptop to perform local (one local MC sweep corresponds to $N$ single spin-flip attempts) and global MCMC, and to train the model, are given in Table~\ref{tab:EA}.

\subsection{Results}
\label{sec:EAres}

We begin by reproducing the results of Ref.~\cite{PhysRevE.101.053312} with the NADE, choosing $N_h=64$ which is the minimal number needed to obtain accurate results according to Ref.~\cite{PhysRevE.101.053312}.
Having reproduced the NADE results of Ref.~\cite{PhysRevE.101.053312}, we also considered a shallow MADE, which has a slightly smaller number of parameters, i.e. $N(N-1)/2 =4950$ versus $2 N N_h=12800$ for $N=100$ and $N_h=64$. The shallow MADE performs identically to the NADE (Fig.~\ref{ener_entro_EA}), and we find the shallow MADE preferable because it has less parameters and it is more interpretable~\cite{trinquier2021efficient}. It might be interesting, in future work, to check whether or not the shallow MADE has a better scaling upon increasing $N$ with respect to the NADE.

Fig.~\ref{ener_entro_EA}d shows the number of local MCMC sweeps  $\tau$ needed to reach a time-averaged spin-spin correlation 
\begin{equation}
C(\tau) = \frac{1}{T}\sum_{t=1}^T \frac{1}N \sum_{i=1}^N \s_i(t)\s_i(t+\tau) = 0.5 \ .
\end{equation}
Note that already for $\beta = 1.5$, the time to decorrelate is of the order of magnitude of $10^4$ sweeps, and it grows quickly upon decreasing temperature. 
We stop investigating the local MCMC at $\beta=1.5$ because below this temperature measuring its relaxation time becomes too computationally costly.

Next, we train the AR models using the sequential tempering procedure with $M=10^5$ training configurations, as detailed in Sec.~\ref{sec:2dEAmethods}. We observe in Fig.~\ref{ener_entro_EA}d that the acceptance rate of global MCMC moves remains very close to one at all temperatures, and it actually {\it increases} upon decreasing temperature, reaching one for $\beta\to\infty$. For this paramagnetic model, new configurations proposed by the global MCMC are completely uncorrelated from previous ones, and
the decorrelation time for the global MCMC is of the order of magnitude of the inverse acceptance rate, hence between one and two global MCMC moves. We checked explicitly in Fig.~\ref{ener_entro_EA}f, where we show that the average Hamming distance (i.e., number of distinct spins) between two configurations obtained by independent global MCMC, and between a single MCMC and itself a few global MCMC later, coincide.

Using the numbers reported in Table~\ref{tab:EA}, we find that at $\beta=1.5$, decorrelation on a standard laptop takes about $0.12$~s using local MCMC (i.e. about $10^{-5}$~s per MC sweep times $12000$ time sweeps to decorrelate), and about $10^{-3}$~s for global MCMC (that requires a few moves to decorrelate, see Fig.~\ref{ener_entro_EA}f, each taking about $3\cdot 10^{-4}$~s), hence with a speedup of two orders of magnitude. Furthermore, the decorrelation time of global MCMC slightly decreases upon lowering temperature, while that of local MCMC strongly increases, leading to an even stronger speedup, divergent for $\beta\to\infty$. Note that the learning time is negligible at large enough $\beta$.

To monitor the quality of the AR models upon decreasing temperature, we compare the exact energy (Fig.~\ref{ener_entro_EA}a) and entropy (Fig.~\ref{ener_entro_EA}b) computed with local MCMC with those estimated by sampling the AR models, finding extremely good agreement.
For the local MCMC computation, we perform an annealing from high temperature measuring the energy and the specific heat, and we obtain the entropy by integrating the latter in temperature. For the AR models, we estimate the energy and the entropy by sampling from the model at each $T$, and computing the average of $H(\s)$ and of $-\log P_{AR}(\s)$ over this sample, respectively.
We note, in particular, that the AR entropy coincides with that of the true model, which shows that the AR model is not mode-collapsed. 
In Fig.~\ref{ener_entro_EA}c, we also show the KL divergences $D(P_{AR}||P_B)$ and $D(P_B||P_{AR})$ as a function of temperature. Both are extremely small, which implies that $\log W(\s) = \log[P_{B}(\s)/P_{AR}(\s)]$ is extremely small for configurations $\s$ that are typical both of the true equilibrium and of the AR model, consistently with the high acceptance rate shown in Fig.~\ref{ener_entro_EA}d (see the discussion in Sec.~\ref{sec:loc_vs_glo_MCMC}).

For future reference, 
we also checked the dependence of the acceptance rate of global MCMC on the training set size $M$. We found that a minimal training set size of about $M=10^4$ is needed to obtain good performance, while below this size the AR model cannot be trained properly and global MCMC fails
(see the SI).

To summarize, we found that:
\begin{itemize}
    \item A 2d EA model with $N=100$ spins can be approximated very accurately by an AR model, as shown by the very small KL divergence.
    \item Correspondingly, global MCMC is very efficient and leads to an important speedup with respect to local MCMC, as shown in Ref.~\cite{PhysRevE.101.053312}. We have shown that the speedup diverges for $\beta\to\infty$.
    \item A simple shallow MADE performs as well as the NADE used in Ref.~\cite{PhysRevE.101.053312}, even with a slightly smaller number of parameters.
    \item A sufficiently large training set ($M\gtrsim 10^4$) must be used to achieve a good efficiency of the global MCMC.
\end{itemize}
We thus conclude that, at least for $N= 100$ (a $10\times 10$ periodic lattice), the 2d EA model is easy to approximate with an AR model, leading to efficient global MCMC. 
Yet, the fact that the AR model concentrates at low temperatures on the two ground states (related by the inversion symmetry), which can be found in short time by algorithms such as McGroundstate~\cite{CJMM22}, leads us to believe that the instance we studied is too small to be really hard to sample.

Furthermore, we repeated the study on a 3d EA model on a $5\times5\times 5$ cubic periodic lattice ($N=125$ spins). Also in this case, because of the small size, the ground state is found easily by McGroundstate~\cite{CJMM22}, despite the fact that the problem is in principle NP-hard~\cite{barahona1982computational}. We found that the AR model approximates well the 3d EA model, leading to efficient global MCMC at all temperatures, despite the presence of a phase transition at finite temperature~\cite{KKY06}, and converging to the ground state in the zero-temperature limit (see the SI).

We believe that a more complete study should investigate systematically the 
dependence on system size for much larger $N$, as well as the dependence on the space dimension, because the EA model is known to become harder and harder to sample upon increasing spatial dimensions. 
Because this is not the main focus of the present work, we leave this for future investigation. The main goal of the present section was to show that our implementation of the global MCMC is able to reproduce and extend previous positive results~\cite{PhysRevE.101.053312}, and it can thus be considered reliable and efficient.


\section{Coloring}
\label{sec:col_res}

We now consider our main benchmark, namely the hard-to-sample coloring problem. We will show that all the global MCMC implementations discussed above fail for this model, and we could not find an implementation that achieves a better efficiency. In particular, we will show that both the maximum-likelihood and the variational training fail, but for quite different reasons.

\subsection{Model}
\label{sec:col_def}

The model is formulated in terms of $N$ variables ${\s_i \in \{0,\cdots,q-1\}}$, each taking $q$ possible colors. The variables are the nodes of an Erd\"os-R\'enyi random graph $\mathcal{G}$, i.e. a graph in which each link $\langle i,j\rangle$ is present at random with the same probability, chosen such that the average connectivity of a node is $c$.
The model
Hamiltonian can then be written as
\begin{equation}
    H(\s) = \sum_{\langle i,j\rangle\in\mathcal{G}} \delta_{\s_i,\s_j} \ ,
\end{equation}
i.e. a $q$-state Potts model where antiferromagnetic couplings (whose value is fixed to one without loss of generality) are only present on the  edges of $\mathcal{G}$.
The Hamiltonian counts the monochromatic edges by assigning a zero energy to all links that connect sites with distinct colors, and energy one to links that connect sites with the same color. Hence, $H(\s)\geq 0$ and $H(\s)=0$ if and only if $\s$ is a proper coloring of $\mathcal{G}$. At zero temperature, the problem of finding the ground state of $H(\s)$ is therefore related to the random graph coloring problem~\cite{mulet2002coloring,Zdeborova2007}. 

\begin{figure}[t]
  \centering
  \includegraphics[width=.95\columnwidth]{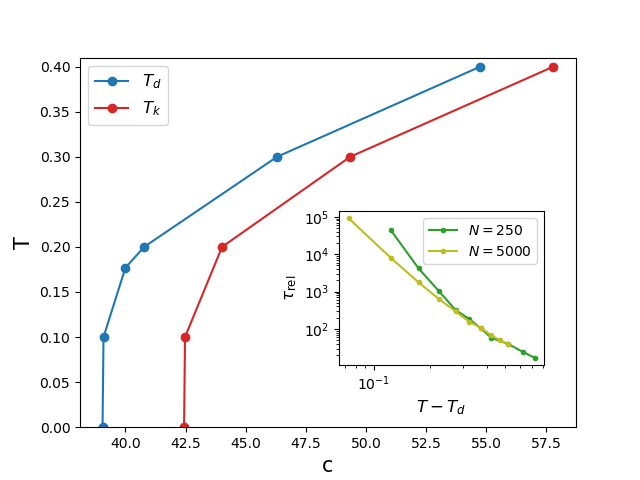}
  \caption{{\bf Coloring -} Phase diagram of the coloring of Erd\"os-R\'enyi random graphs with $q=10$ colors, as a function of the average node connectivity $c$ and of temperature $T$~\cite{cavaliere2021optimization}. The upper line is the dynamical glass transition $T_d(c)$, below which the decorrelation time increases exponentially with $N$; the lower line is the condensation transition $T_K(c)$ below which quiet planting is not possible.
  Inset: decorrelation time of the equilibrium local MCMC for $c=40$, for two different system sizes, as a function of $T-T_d(c=40)$.
  }
  \label{phasediag_q10}
\end{figure}

At finite temperature and for large enough $q$, the model belongs to the Random First Order Transition (RFOT) universality class~\cite{KZ08}, together with many other optimization problems~\cite{KMRSZ07} and with mean-field structural glasses~\cite{KW87,KT88,CK93,parisi2020theory}.
These models are characterized by a finite-temperature {\it dynamical glass transition}, below which the time needed for a proper sampling of the Boltzmann probability diverges exponentially with $N$~\cite{MS06}, due to the presence of an exponential number of peaks (i.e. metastable states)~\cite{CK93,monasson1999determining,KMRSZ07,parisi2020theory}.

The coloring problem is specified by the number of sites $N$, the choice of the random graph $\mathcal{G}$ which is specified by its average connectivity $c$, the number of colors $q$, and temperature $T$.
For the model to be a good representative of the RFOT class, we need $q$ to be large enough, otherwise the transition is too close to a second order one, especially at finite $N$~\cite{KZ08}. We thus chose $q=10$, for which
a finite RFOT dynamical transition $T_d(c)$ is present for all $c>39.02$, and the corresponding phase diagram is reported in Fig.~\ref{phasediag_q10} (data were privately communicated by the authors of Ref.~\cite{cavaliere2021optimization}).

Systems in the RFOT class also display a second phase transition, called condensation or Kauzmann transition $T_K(c)$. Below this transition, the number of metastable glassy states become sub-extensive~\cite{KMRSZ07}.
For our purposes, this phase transition is relevant because for $T>T_K(c)$ one can use the {\it quiet planting} technique~\cite{ZK16}, which consists in exchanging the sampling of $\s$ and the generation of~$\mathcal{G}$. While one should first generate a graph $\mathcal{G}$ and then sample $\s$ from the Boltzmann distribution for that~$\mathcal{G}$ (which is exponentially hard in $N$ at low $T$), in quiet planting one first takes a random configuration $\s$, and then constructs at a given $T$ a random graph $\mathcal{G}$ such that $\s$ is an equilibrium configuration for the model defined on that graph, at temperature $T$ (which can be done in polynomial time). It can be shown~\cite{ZK16} that for $T>T_K(c)$ in models such as the coloring, this exchange of order does not affect the statistical properties of $\mathcal{G}$.
This allows one to run local MCMC from the configuration $\s$, which is guaranteed to be an equilibrium one, thus bypassing the need for equilibration, which takes an exponentially large time for $T<T_d(c)$.
Our choice of $q=10$ ensures the existence of a large enough region $T_K<T<T_d$ where model-generated samples may be compared to the planted configurations.

\begin{figure*}[t]
    \centering
    \includegraphics[width=\textwidth]{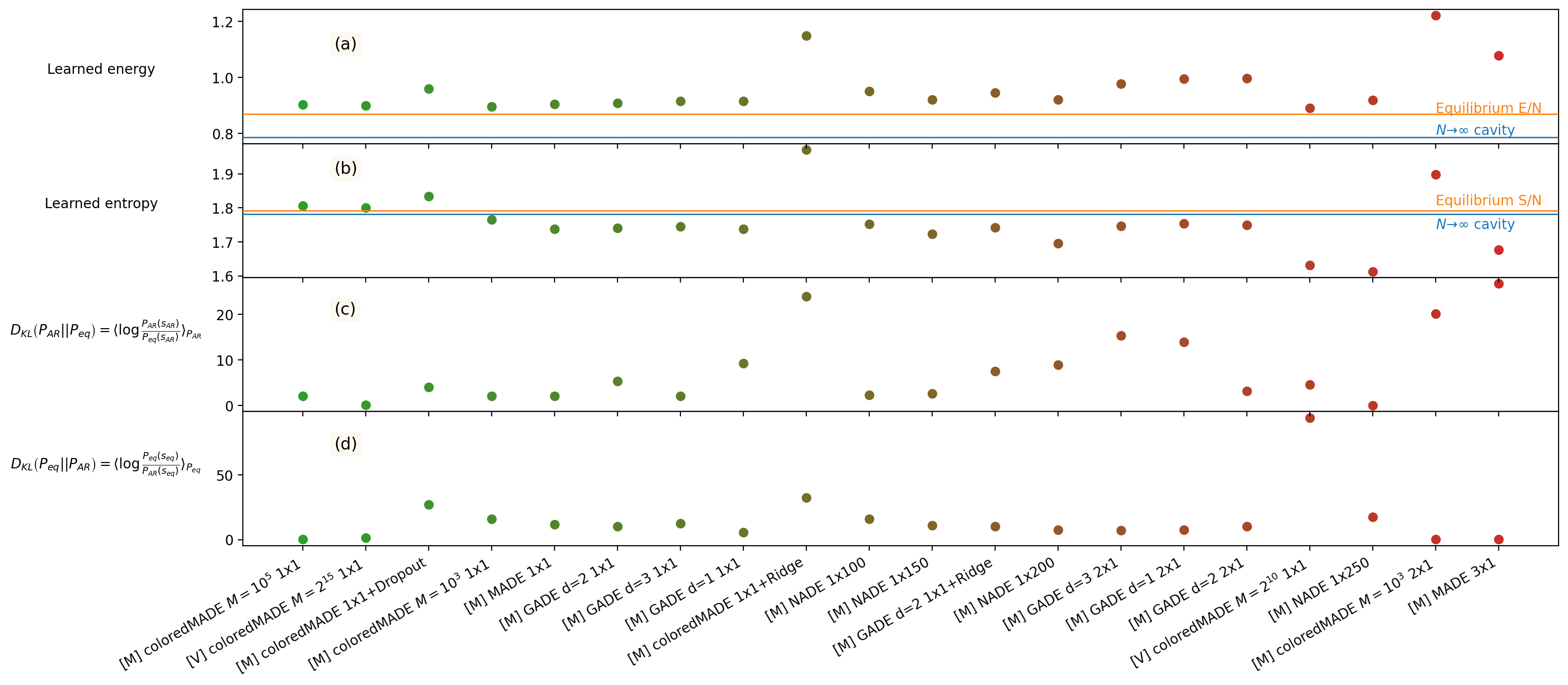}
    \caption{
        {\bf Coloring} with $q=10$, $c=40$, $T=1$ and $N=250$.
        The four panels of the figure report the value of energy (a), entropy (b) and KL divergences (c, d) for different AR architectures, trained with the techniques of Sec.~\ref{sec:train}.
    The name of the models (i.e. its training technique, AR network type, etc.) is reported on the x-axis and corresponds to a specific color.
    Overall, the models are ranked, from left to right, by free energy, which at $T=1$ is the difference of energy minus entropy.
    The horizontal lines correspond to the equilibrium values, measured using local MCMC for $N=250$ (orange, the entropy is obtained via thermodynamic integration) or calculated from the cavity method for $N\to\infty$ (blue).
    }
    \label{fig:learnedE_b1}
\end{figure*}
For our study, we thus chose $c=40$, for which ${T_d= 0.1768}$ and $T_K=0$ (Fig.~\ref{phasediag_q10}). In this way, we can perform quiet planting at any temperature to study the equilibrium local MCMC dynamics, and we also have a wide range of temperature $T<T_d$ where this dynamics is exponentially slow in $N$.
Notice that such graphs are full of short loops, so some finite size effects are measurable, like an energy drift in the planted solution.
We have confirmed that the scenario we describe in the next sections is equally reproduced for lower connectivity, $\{q=5,c=13\}$ and $\{q=10,c=30\}$, so that it is mainly determined by the complexity of the problem. 
In Fig.~\ref{phasediag_q10} we show the decorrelation time of the overlap correlation function,
\begin{equation}\label{eq:COLove}
C(t,\tau) = \frac{1}N \sum_{i=1}^N \delta_{\s_i(t),\s_i(t+\tau)} \ ,
\end{equation}
here measured using local MCMC and averaging over time~$t$. 
Note that with this definition, complete decorrelation corresponds to $C(t,\tau) \sim 1/q$.
The decorrelation time is defined by ${\langle C(t,\tau) \rangle_t=0.5}$. When plotted
as a function of $T-T_d$, it displays the characteristic power-law divergence predicted by RFOT theory. In the vicinity and below $T_d$, the dynamics becomes so slow that we are unable to observe decorrelation of local MCMC in reasonable time. We consider a fixed size $N=250$, which is a good compromise between avoiding too large finite size effects, and having a small enough size to allow for an extensive testing of autoregressive model architectures.

It is also important to mention that the thermodynamics of the model can be solved, for $T>T_K(c)$ and for $N\to\infty$, by a simple cavity computation~\cite{KZ08,ZK16,cavaliere2021optimization}.
In particular, the energy and entropy per spin are given by
\begin{equation}\label{eq:cavity}
\begin{split}
    e(T) &= \frac{c}2 \frac{e^{-\beta}}{q-1+e^{-\beta}} \ , \\
    s(T) &= \log q + \frac{c}2 \log\left[ \frac{q-1+e^{-\beta}}{q}\right] + \beta e(T) \ .
\end{split}\end{equation}
Although for $N=250$ we observe significant deviations from the thermodynamic values (Fig.~\ref{fig:learnedE_b1}), these expressions are still useful as a reference.

\subsection{Model selection}

Our goal is to use autoregressive models to generate good trial configurations and speed up the dynamics of the hard-to-sample coloring problem.
Before exploring the hard region below $T_d$, we test how different AR architectures perform in the paramagnetic phase at $T=1$, where the local MCMC dynamics is fast and we can easily generate an equilibrium sample from the Boltzmann distribution. Our intermediate goal is to select the best architectures and training methods (see the SI for details).

For the AR decomposition in Eq.~\eqref{eq:BayesAR}, variables are ordered naively, which means that we first fix an arbitrary ordering $\sigma_1,\cdots,\s_N$, and then construct a graph $\mathcal{G}$ as described in Sec.~\ref{sec:col_def}. We also considered ordering the variables from higher to lower connectivity, and found no qualitative difference. Results are reported here for this second choice of ordering.
The models are trained either via maximum-likelihood on a sample of $M$ equilibrium configurations generated via local MCMC, or via the variational method using $M$ model-generated samples to compute the gradient, see Sec.~\ref{sec:var}. 
The training is done using the Adam optimizer~\cite{kingma16} and mini-batches of $1024$ configurations. In order to enforce color symmetry, each configuration in the sample is subjected to a random permutation of colors before being used, in each epoch of training.
We start from a learning rate of $0.01$ that we reduce by a factor $2$ after every $250$ epochs of no improvement~\cite{bengio12}.
We set an early stop criterion that triggers if the likelihood of a validation set made of independent equilibrium samples starts decreasing. 

\begin{figure*}[t]
    \centering
    \includegraphics[width=.8\textwidth]{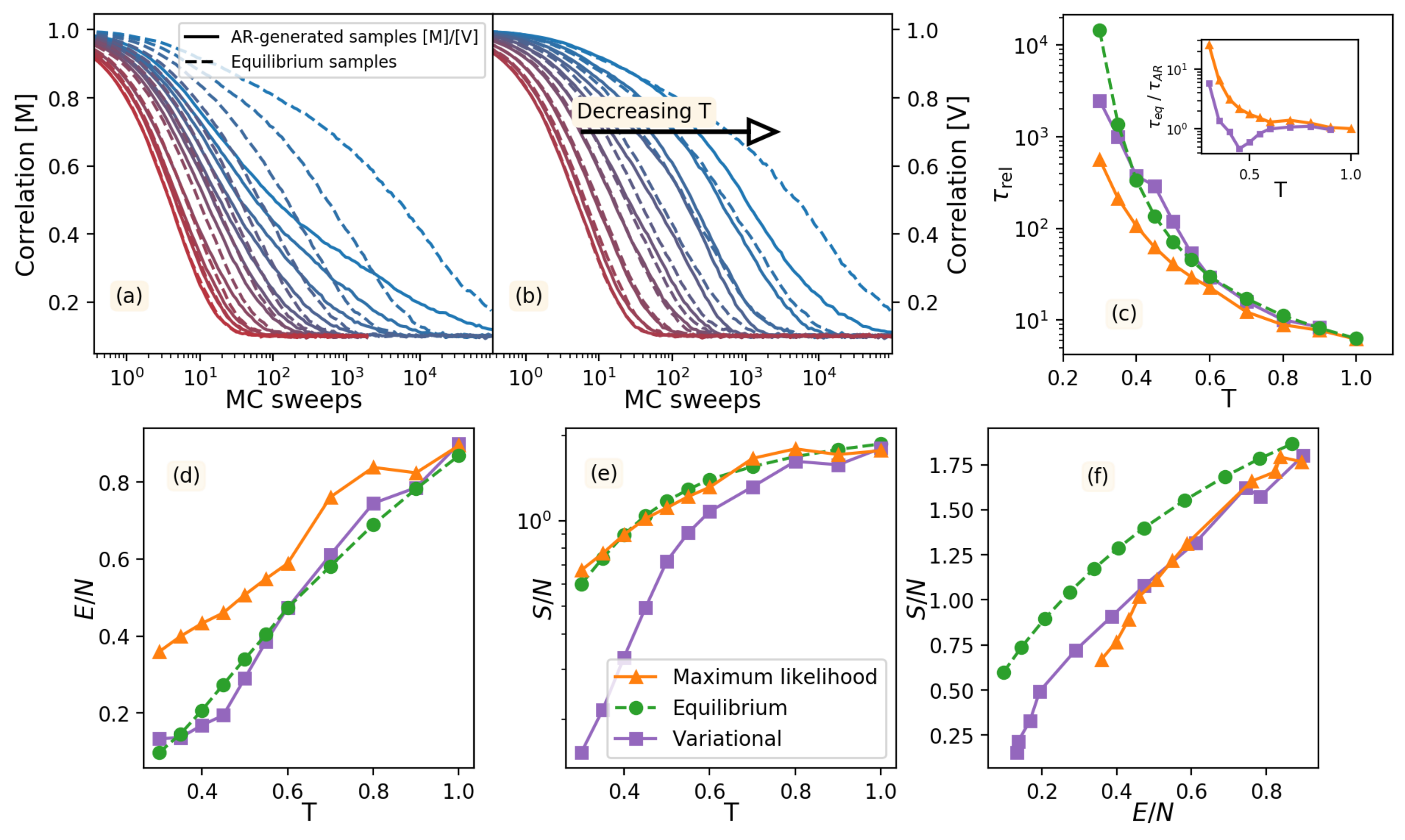}
    \caption{
    {\bf Coloring} with $q=10$, $c=40$ and $N=250$.
    (a,b) Overlap correlation function for {\it local} MCMC dynamics starting from samples generated either in equilibrium (dashed lines, via long enough local MCMC) or by a ColoredMADE AR model trained both by maximum likelihood (a, continuous lines) and variationally (b, continuous lines). The temperature decreases from $T=1$ (red) to $T=0.3$ (blue) following the color gradient.
    (c) Relaxation time $\tau_\mathrm{rel}$ defined from $C(0,\tau_\mathrm{rel})=1/e$ for the samples generated by the AR model trained using maximum likelihood (orange) or variationally (purple) compared to equilibrium (green).
Relaxation is measured averaging over $100$ local MCMC runs.
In the inset we report the ratio of the equilibrium time to that of AR models.
The difference becomes larger upon decreasing the temperature, but the variational model (purple) remains accurate down to $T\sim0.35$.
(d) Average energy and (e) entropy of samples generated by the two AR models and in equilibrium, with the same color code as in (c). In panel (e), the green data correspond to the exact entropy in the thermodynamic limit, Eq.~\eqref{eq:cavity}. Using the same data of (d)-(e), we plot in (f) the entropy of the generated samples as a function of their energy.  
}
    \label{fig:gensamples}
\end{figure*}

At the end of the training, we compute the model energy (Fig.~\ref{fig:learnedE_b1}a) and entropy (Fig.~\ref{fig:learnedE_b1}b) by sampling from the AR model and computing the average of $H(\s)$ and $-\log P_{AR}(\s)$, respectively.
We compare these results with the exact ones obtained via local MCMC. 
We observe that the AR energy and entropy are generally quite close to the equilibrium values. 
In particular, the samples generated by the models often have an energy that is closer to the average equilibrium energy than that of the planted configuration, hence the difference falls within the typical equilibrium fluctuations.
As a general trend, we observe that increasing the model expressivity does not improve the energy, probably because of overfitting, as shown by the decrease in entropy. On the other hand, adding a regularization (either dropout or ridge) degrades the performance significantly, by increasing both the energy and the entropy. We thus conclude that the best model is once again a shallow MADE (ColoredMADE, see the SI), which can achieve a very low energy and a very high entropy, close to the target values, both with variational and maximum-likelihood training.

In Fig.~\ref{fig:learnedE_b1}c,d we report the KL divergences between the Boltzmann distribution and that of the AR models, which confirms that the two shallow ColoredMADE have the smallest KL divergence.
Having selected two good architectures at $T=1$, we now focus on these two models and
 check how they perform upon lowering the temperature.
We consider, for each $T$, a ColoredMADE trained by maximum likelihood on an independent equilibrium sample generated from the Boltzmann distribution by long enough local MCMC at that $T$, and another ColoredMADE trained variationally at fixed $T$ (which, we recall, does not require to generate equilibrium samples). 

We will show next that both models fail at lower $T$.
Of course, there might be other architectures that we did not consider here, which might provide better results. However, because both adding regularization or increasing the expressivity of the model (also by using GNNs) did not seem to improve the efficiency, we do not see a clear path for improvement.

\subsection{Failure of sequential tempering}
\label{sec:MLfail}

We first consider the ColoredMADE trained by maximum likelihood, and
compare the average energy of the equilibrium configurations with those generated by the AR model (Fig.~\ref{fig:gensamples}d). We observe that the model generates configurations with higher energy as soon as temperature goes below $T\sim 1$, while its entropy
(Fig.~\ref{fig:gensamples}e) remains close to the equilibrium value. Yet, once the entropy is plotted parametrically as a function of energy (Fig.~\ref{fig:gensamples}f), the difference from the equilibrium result becomes apparent.

\begin{figure*}[t]
  \centering
  \begin{minipage}{.7\textwidth}
    \includegraphics[width=\textwidth]{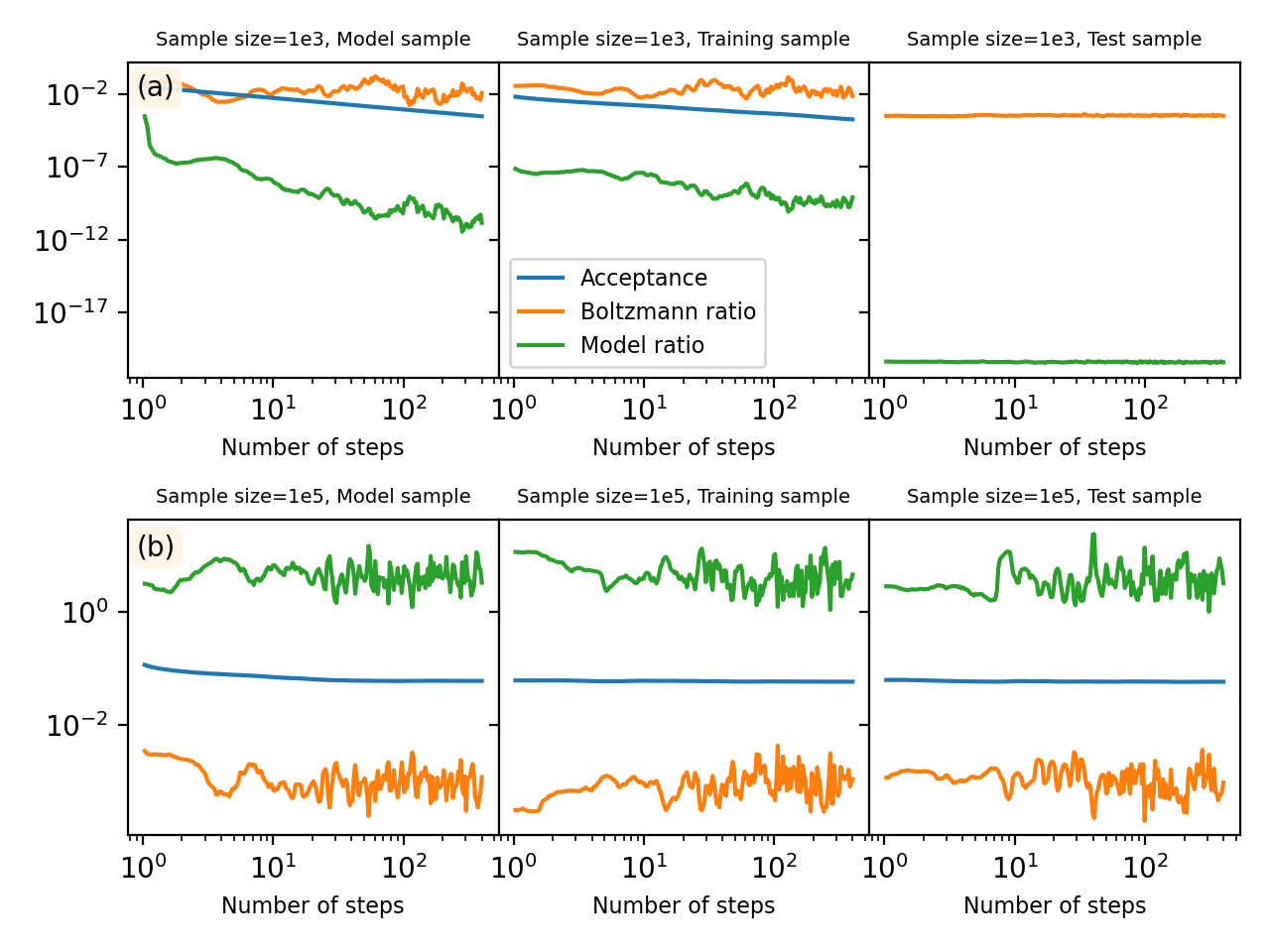} 
    \end{minipage}
    \begin{minipage}{.29\textwidth}
  \includegraphics[width=\textwidth]{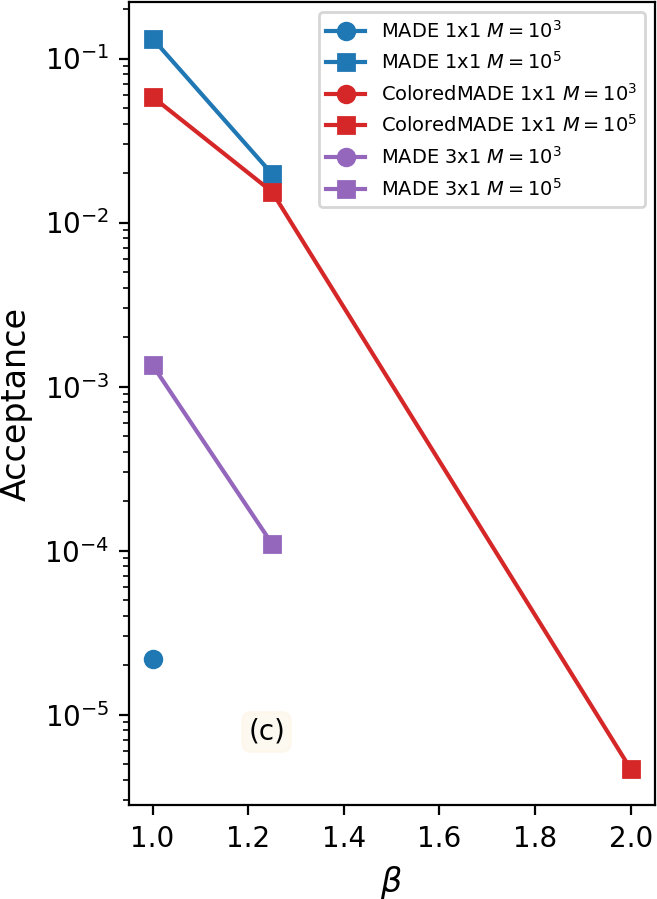} 
  \end{minipage}
  \caption{{\bf Coloring} with $q=10$, $c=40$ and $N=250$. (a) Acceptance rate of global MCMC for the ColoredMADE trained via maximum likelihood at $T=1$ using a sample of $M=10^3$ equilibrium configurations. 
  Here, $K=100$ independent global MCMC are initialized in configurations generated by the AR model itself (model sample), taken from the training sample, or from additional independent equilibrium configurations (test sample). We report the average acceptance rate, and the geometric average of the Boltzmann and model ratio, over the $K$ chains.
  (b) Same with $M=10^5$. 
 (c) Acceptance rate (on the test set) of global MCMC for several AR models trained with the maximum likelihood technique on equilibrium configurations generated via local MCMC at several temperatures.  Missing data indicate that the acceptance rate is so small that it is impossible to measure it.
}
  \label{fig:acceptance_coloring_time}
\end{figure*}

We also compare the correlation functions of local MCMC dynamics starting from the equilibrium samples  and the AR-generated samples. (To avoid confusion, we stress that here we are not yet considering AR-assisted global MCMC.)
In Fig.~\ref{fig:gensamples}a we report the overlap correlation function defined by Eq.~\eqref{eq:COLove}, here for fixed $t=0$ as a function of $\tau$ measured in MC sweeps. 
The dashed lines are obtained starting at $t=0$ from equilibrium samples, while the full lines are obtained starting from samples generated by the AR model.
We observe that AR-generated samples have the correct equilibrium local MCMC dynamics only at very high temperatures. In Fig.~\ref{fig:gensamples}c we compare the relaxation times, defined by $C(0,\tau_{\rm rel})=1/e$, and show that AR-samples have systematically faster dynamics upon lowering $T$.
Overall, these results suggest that the maximum-likelihood training is effective at high temperatures only, because upon lowering temperature the model is unable to propose configurations of low enough energy. This study is restricted to $T\geq 0.3$ because below that temperature we are unable to generate the equilibrium samples needed to train the model by local MCMC.

Next, we consider the global MCMC dynamics, and we measure its acceptance rate at $T=1$.
As in Sec.~\ref{sec:EAres}, global MCMC can be initialized (i) on AR-generated configurations, (ii) on the training set and (iii)~on an independent test set made by additional equilibrium configurations.
In Fig.~\ref{fig:acceptance_coloring_time}a, we consider a ColoredMADE trained on ${M=10^3}$ configurations, and
we report the average acceptance rate together with the geometric average of the Boltzmann ratio $P_B(\s_{new})/P_B(\s_{old})$ and the model ratio $P_{AR}(\s_{old})/ P_{AR}(\s_{new})$.
Similarly to the 2d EA model (see the SI), we observe that for such a small $M$, the acceptance rate on the test set is vanishingly small, while using the other initializations, it decreases with time. The situation improves strongly upon increasing the size of the training set to $M=10^5$ (Fig.~\ref{fig:acceptance_coloring_time}b), leading to a good acceptance rate. Global MCMC can then decorrelate reasonably fast in this case.

We explore more systematically the dependence on $T$ and $M$ in Fig.~\ref{fig:acceptance_coloring_time}c, where we report the acceptance as a function of $\beta$ for different architectures and different $M$. 
We observe that for a given architecture and $M$, the acceptance drops exponentially fast with $\beta$, in some cases becoming so small that we cannot measure it (no move is accepted during the longest global MCMC we can run).
We conclude that the maximum likelihood approach fails because, upon lowering temperature, the AR models are not expressive enough to properly fit the target energy, and as a consequence they propose moves that cannot be accepted by the global MCMC. Increasing $M$ does not help at low temperature (Fig.~\ref{fig:acceptance_coloring_time}c), and we have already seen that increasing the model expressivity only leads to overfitting (Fig.~\ref{fig:learnedE_b1}).

It is possible that by improving model expressivity and at the same time considering larger $M$, the acceptance rate could improve. However, the AR architectures we considered are already quite hard to train, and $M\sim10^5$ corresponds to a quite large dataset, and still the method fails at $T=0.5$, which is a quite high temperature in the paramagnetic phase. These results thus highlight the lack of computational efficiency of this approach. 

Note that the global MCMC fails even when the AR model is trained using maximum likelihood on a perfect equilibrium sample, i.e. under the best possible conditions. Hence, sequential tempering cannot work, because upon decreasing temperature the global MCMC remains stuck and does not lead to a proper new sample on which the AR model can be trained. In fact, performing sequential tempering runs (not shown) we observe that the global MCMC does not equilibrate, hence the AR model has even worse energy and leads to an even slower global MCMC.

\subsection{Failure of the variational training}
\label{sec:VARfail}

\begin{figure}[t]
  \centering
    \includegraphics[width=\columnwidth]{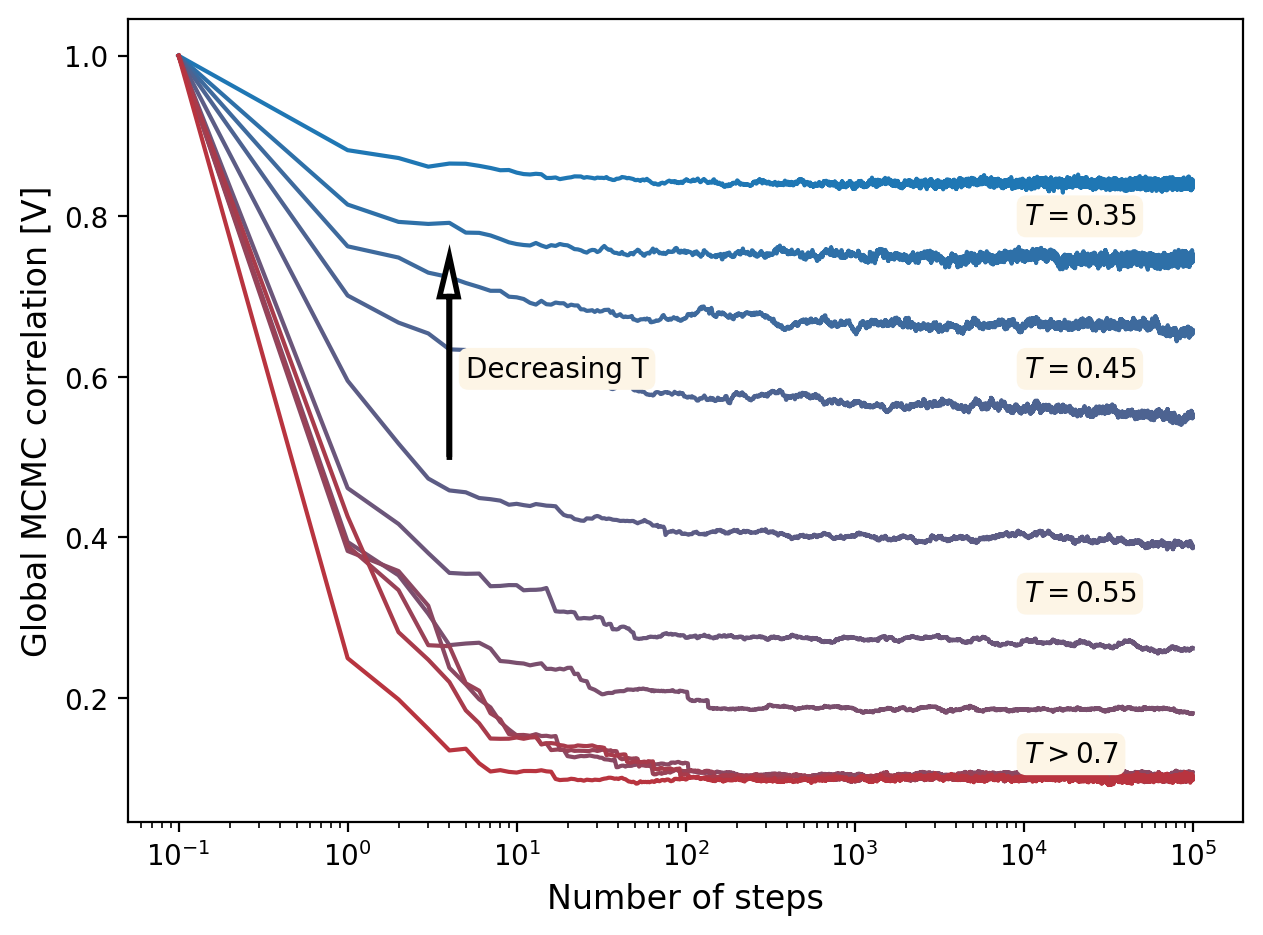} 
\caption{{\bf Coloring} with $q=10$, $c=40$ and $N=250$.  Overlap correlation function of the global MCMC using the variational ColoredMADE as a function of the number of MC steps.
}
  \label{fig:entropy}
\end{figure}

\begin{figure}[t]
  \centering
    \includegraphics[width=\columnwidth]{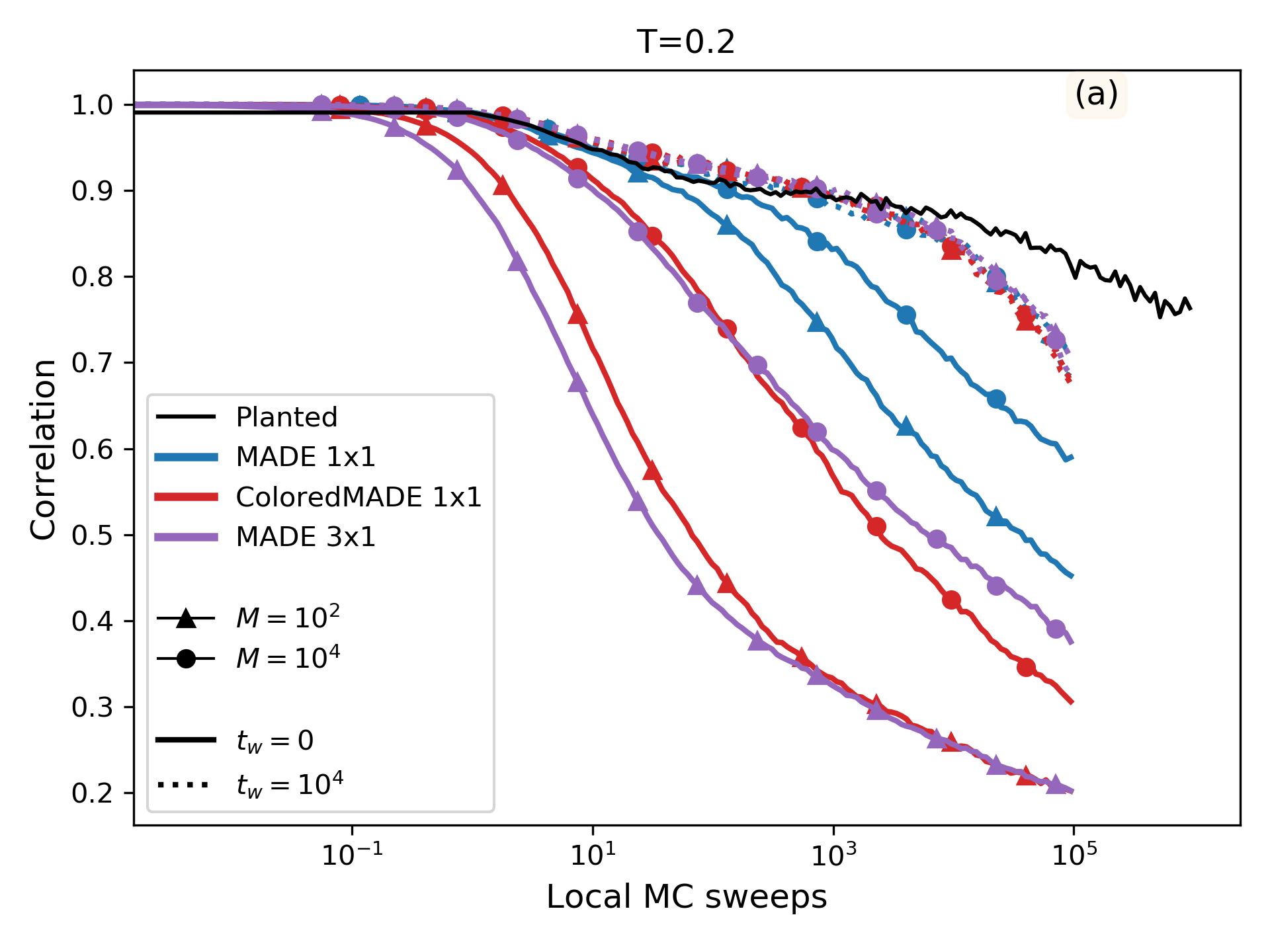} 
     \includegraphics[width=\columnwidth]{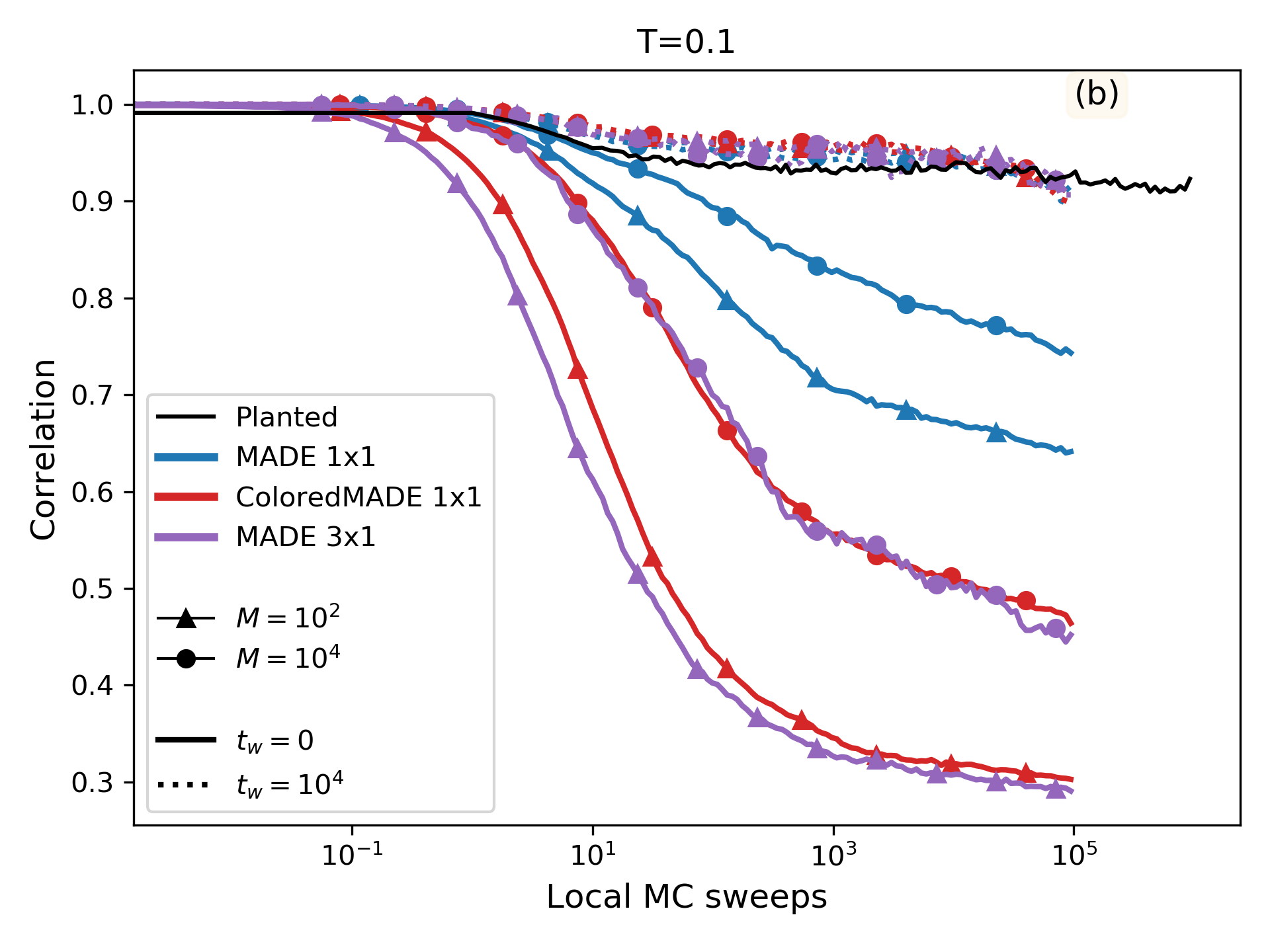} 
  \caption{{\bf Coloring} with $q=10$, $c=40$ and $N=250$.  
Correlation $C(t_w,\tau)$ of local MCMC initialized at $t=0$ in AR-generated samples 
at low temperatures $T=0.2$ (a) and $T=0.1$ (b).
We used the following model architectures: shallow MADE (blue), shallow ColoredMADE (red) and deep MADE ($D=3$, purple). The models have been trained via the variational approach with batches of $M=10^2$ (triangles) or $M=10^4$ (circles) configurations. The full line is the correlation measured from $t_w=0$, while the dashed line is measured after waiting $t_w=10^4$ sweeps. 
In black we report, for comparison, the exact equilibrium dynamics, obtained initializing the local MCMC in the planted configuration. 
}
  \label{fig:var}
\end{figure}

We now consider the variationally-trained ColoredMADE, and note that its
energy remains accurate down to $T=0.3$
(Fig.~\ref{fig:gensamples}d), while its entropy decreases quickly upon lowering $T$ (Fig.~\ref{fig:gensamples}e), suggesting mode collapse, i.e. the model only learns a limited portion of the support of the Boltzmann distribution. Surprisingly, the parametric curve of entropy versus energy (Fig.~\ref{fig:gensamples}f) is very similar to that obtained using maximum-likelihood training.
Consistently, when we run {\it global} MCMC using the variational model, we observe that the acceptance rate remains good, but the proposed moves do not lead to any decorrelation (Fig.~\ref{fig:entropy}), and as a result the sampling of the Boltzmann distribution is not efficient. Global MCMC indeed remain confined in the limited portion of the phase space that has been learned by the variational model. This is consistent with the discussion of Sec.~\ref{sec:CWmodecollapse} for the CW model.

In Fig.~\ref{fig:gensamples}b, we report the correlation functions of the {\it local} MCMC dynamics initialized both in equilibrium and in configurations generated by the variational AR model, and note that they are essentially indistinguishable down to $T\sim 0.35$.
 Because this method does not require the generation of equilibrium samples for training,
one might wonder whether, despite being mode-collapsed and thus useless for {\it global} MCMC, the variational model could still generate configurations that are close enough to some of the equilibrium ones, such that the {\it local} MCMC dynamics coincides with the equilibrium one. If true, this would be interesting, because one could use the variational model to generate {\it some} typical equilibrium states, although only a small subset of them. However, we show in Fig.~\ref{fig:var} that this is not the case. Below $T\sim 0.3$, upon approaching the glass transition, the energy of the variational model remains higher than the equilibrium one, and the generated configurations display an initially faster local MCMC dynamics, whose relaxation time increases during the evolution (the so-called aging behavior typical of glasses). Hence, the variational model also fails in providing good equilibrium configurations.

\subsection{Role of the color symmetry}

One might wonder whether the failure of the AR models is somehow related to the presence of the color symmetry (see the SI for details), which might lead to a proliferation of states, thus requiring a lot of training samples to be taken into account correctly. Or, maybe, breaking the color symmetry would allow for a better learning of the AR model, that could use local fields to obtain more information on the Boltzmann distribution, thus reducing the role of couplings.

To address this question, we tried to break the color symmetry by adding small random local fields on each spin, thus lifting the degeneracy of the   $q!$ equivalent configurations. Recall that $q=10$, hence $q!$ is very large, although the contribution to the entropy per spin, $\log(q!)/N$, is negligible.
Yet, we repeated the same analysis and we obtained very similar results.
We thus conclude that the addition of small external fields is not enough to make the approach work, because its failure is due to the intrinsic difficulty of representing the multiplicity of states typical of a glassy system.

\section{Conclusions}

In this paper, we addressed the general problem of whether machine learning can assist Monte Carlo methods to provide a computational speed up in hard-to-sample glassy problems. 

We confirmed previous findings~\cite{Wu2019,PhysRevE.101.053312,gabrie2021adaptive,wu21,Hibat-Allah2021,fan2021finding,schuetz2022graph} that identified a class of problems, such as simple ferromagnetic models and the two-dimensional Edwards-Anderson spin glass model, which can be well approximated by simple autoregressive models. These models can then be used to sample the Boltzmann distribution by machine-learning-assisted global MCMC, providing a computational speedup of several orders of magnitude.
For these type of problems, several different architectures and training schemes have been proposed, but we found that all these methods are roughly equivalent. In particular, we tend to prefer a shallow autoregressive architecture, which provides a simple and interpretable model.
We also highlighted the need for a large enough training sample for the efficiency of the approach.

We found that
the situation is totally different when one considers really hard-to-sample problems belonging to the Random First Order Transition (RFOT) universality class, such as the
coloring of random graphs and, possibly, the model considered in Ref.~\cite{condmat7020038}. For these type of problems, the landscape is very rough and complex, displaying a multitude of locally stable glass states that are known to trap standard local MCMC~\cite{MS06}. We showed that machine-learning-assisted global MCMC fails, either because of the mode-collapse of the variationally learned model into a restricted portion of the phase space (that is not even representative of equilibrium in the vicinity of the glass transition), or because the sequential tempering scheme fails to achieve a high enough acceptance rate of global MCMC moves.

We tried several architectures in order to overcome the problem, but we did not find any promising sign of improvement. In particular, we found that adding regularization leads to underfitting (high entropy), while increasing the expressivity of the model leads to overfitting (low entropy), and both are detrimental to the global MCMC efficiency.

We are thus led to believe that there is an intrinsic difficulty of these problems that prevents machine learning methods to solve them efficiently. Of course, we might have missed something and future work could find a solution using smarter architectures. Yet, we believe that our work is useful because it establishes a benchmark against which, in our opinion, such machine learning methods should be tested to assess their performance as universal methods for sampling speedup.

\acknowledgments

We thank
Giulio Biroli, Giuseppe Carleo, Angelo Cavaliere, Marylou Gabri\'e, Ilya Nemenman and Federico Ricci-Tersenghi for many useful suggestions related to this work.

This project has received funding from the European Research Council (ERC) under the European Union's Horizon 2020 research and innovation programme (grant agreement n. 723955 - GlassUniversality). 
J.T. is supported by a PhD Fellowship of the i-Bio Initiative from the Idex Sorbonne University Alliance.



%

\clearpage
\newcommand{\beginsupplement}{
        \clearpage
        \setcounter{section}{0}
        \renewcommand{\thesection}{S\arabic{section}}
        \setcounter{equation}{0}
        \renewcommand{\theequation}{S\arabic{equation}}
        \setcounter{table}{0}
        \renewcommand{\thetable}{S\arabic{table}}
        \setcounter{figure}{0}
        \renewcommand{\thefigure}{S\arabic{figure}}
     }

\beginsupplement

\section{Architectures}
\label{sec:architect}

In this section we give the details of the different architectures that are used in the paper.

\subsection{Representation of the input variables}
\label{sec:non-bin}

Because we deal with autoregressive models taking as input $N$ variables $\s=\{\s_1,\cdots, \s_N\}$, we define a notation $\s_{<i} = \{\s_1,\cdots,\s_{i-1},0,0,\cdots,0\}$, which is a $N$-dimensional vector in which the variables with index $\geq i$ are ``masked'' to zero. To avoid confusion, let us stress that $\s$ is a $N$-dimensional vector, $\s_i$ are its components, and $\s_{<i}$ is a masked $N$-dimensional vector.

For non-binary variables $\s_i \in \{ 0,1,\cdots,q-1 \}$ that can assume $q>2$ states (or ``colors''), the input values can be efficiently represented as binary vectors via one-hot encoding.
The standard one-hot-encoding recipe consists in replacing $q$-state variables by binary $q$-component vectors $\hat \s_i$, in which a single component equal to one represents the state, e.g., $\s_i=2\xrightarrow[]{}\hat \s_i= (0,0,1,0,...,0)$.
More generally, $\hat \s_i^k = \delta_{k,\s_i}$, for $k=0,\cdots,q-1$.

If we use a one-hot-encoding representation of non-binary variables, we define 
$\hat \s_{<i} = \{\hat \s_1, \cdots, \hat \s_{i-1}, \hat 0, \cdots, \hat 0 \}$ as a $(N\times q)$-component vector in which we set all the vectors with 
index $\geq i$ to have all components equal to zero.

\subsection{Shallow model}
\label{sec:shallow}

The shallow model is the simplest model that satisfies the autoregressive property (beyond independent variables). 
In this model, each conditional probability is written in the form of a Boltzmann distribution with local fields and two-variable couplings. For binary variables $\s_i\in \{0,1\}$, it is conveniently parametrized as:
\begin{equation}\label{eq:ARbinary}
    P_{AR}^i(\s_i|\s_{<i}) = \frac{ \exp \left( \sum_{j(<i)}J_{ij}\s_i\s_j+h_i\s_i \right)
    }{2 \cosh \left( \sum_{j(<i)}J_{ij}\s_j+h_i \right)
    } \ ,
\end{equation}
such that the total number of parameters of the AR model is $N(N-1)/2+N$.

For non-binary variables, 
using a one-hot-encoding representation (Sec.~\ref{sec:non-bin}),
the shallow model reads:
\begin{equation}\label{eq:one-hot-shallow}
    P_{AR}^{i}(\hat \s_i|\hat \s_{<i}) =\frac{ \exp \left( \sum_{j(<i)}\hat \s_i^T \cdot {\rm J}_{ij} \hat \s_j+\hat \s_i^T \cdot \hat h_i \right)}{ \sum_{\s=0}^{q-1} \exp \left( \sum_{j(<i)}\hat \s \cdot {\rm J}_{ij}\hat \s_j+\hat \s \cdot \hat h_i \right)} 
    \ ,
\end{equation}
where the ${\rm J}_{ij}$ are $q\times q$ real (non-symmetric) matrices, the $\hat h_i$ are real $q$-component vectors, and the AR model has in total $q^2N(N-1)/2+Nq$ parameters. 

Note that there is an overparametrization in Eq.~\eqref{eq:one-hot-shallow}, because for instance one of the $q$ components of $\hat h_i$ can be set to zero without loss of generality, due to the normalization factor. Similarly, one line and one column of the matrix ${\rm J}_{ij}$ can be set to zero. This so-called ``gauge invariance'' slightly reduces the number of parameters, such that in particular for $q=2$ one recovers the binary representation in Eq.~\eqref{eq:ARbinary}.

\subsection{Color symmetry}
\label{sec:csym}

In this paper, we will consider in particular the coloring problem, which enjoys a `color' symmetry, meaning that the Hamiltonian is invariant under any arbitrary permutation of the $q$ colors. We then find useful, in order to (partially) prevent mode-collapse, to enforce the same symmetry into the AR model. 
This requires in particular the vector $\hat h_i = h_i (1,\cdots,1)$ to be a constant, hence the term $\hat \s_i^T \cdot (h_i,\cdots,h_i) = h_i$ becomes a constant too.
Furthermore, we note that the only $q\times q$ matrices that are fully invariant under permutations of the $q$ states are the identity matrix and the matrix of all ones. The identity matrix with a number $J_{ij}$ on the diagonal gives a contribution $J_{ij} \hat \s_i^T \cdot \hat \s_j = J_{ij} \delta_{\s_i,\s_j}$, while the matrix of all ones gives again a constant contribution. Constant terms can be neglected due to the normalization, and the model becomes
\begin{equation}\label{eq:PARsym}
    P_{AR}^{i}( \s_i|\s_{<i}) =\frac{ \exp \left( \sum_{j(<i)} J_{ij} \delta_{\s_i,\s_j}  \right)}{ \sum_{\s=0}^{q-1} \exp \left( \sum_{j(<i)} J_{ij} \delta_{\s,\s_j} \right)} \ ,
\end{equation}
which reduces the number of parameters to $N(N-1)/2$, i.e. by a factor $\sim q^2$.

Note that, introducing a $q$-component probability vector $\hat P^i_{AR} = \{P^i_{AR}(\s_i=0),\cdots,P^i_{AR}(\s_i=q-1)\}$, Eq.~\eqref{eq:PARsym} can be written as
\begin{equation}\label{eq:PARsymvec}
\begin{split}
\hat P^i_{AR} &=    \text{softmax}(H_i^0,\cdots,H_i^{q-1})
    \ , \\
    H_i^k &= \sum_{j(<i)} J_{ij} \delta_{k,\s_j} =  \mathcal{F}_i[\s^k] \ ,
\end{split}
\end{equation}
i.e. the probability of $\s_i=k$ is the softmax (over $k$, at fixed $i$) of a vector $H_i^k$. Each of the $N$-dimensional vectors $H^k = \mathcal{F}[\s^k]$ is calculated independently by applying a function $\mathcal{F}$ {\it with the same weights} $J_{ij}$ to the $N$-dimensional one-hot-encoded input vector $(\s^k)_i=\delta_{k,\s_i}$. The function $\mathcal{F}$ satisfies the AR property, hence its component $i$ only depends on the components with $j<i$ of the input.

While in Eq.~\eqref{eq:PARsymvec} the function $\mathcal{F}$ is a simple linear layer (with the AR mask), more general architectures can be considered while keeping the color symmetry, by replacing $\mathcal{F}$ by an arbitrarily complex neural network with multiple hidden layers and non-linearities. The procedure can be summarized as follows:
\begin{enumerate}
    \item Construct the $q$ one-hot-encodings of the input, $(\s^k)_i = \delta_{k,\s_i}$, each $\s^k$ being a binary vector of size $N$ (i.e. the number of spins).
\item Apply {\it the same} (i.e. independent of $k$) arbitrarily complex neural network $\mathcal{F}$ to each of the $q$ inputs to construct $q$ output fields also of size $N$, i.e.
\begin{equation}
    H^k = \mathcal{F}[\s^k] \ ,
\end{equation}
where the function $y = \mathcal{F}[x]$ satisfies the AR property, such that $y_i$ only depends on $\{x_j\}_{j<i}$.
\item Take a softmax of the $q$-component field $(H_i^0,\cdots,H_i^{q-1})$ to obtain the probability of variable~$\s_i$.
\end{enumerate}
Because the same neural network  with the same weights is applied independently  for all $k$, this process guarantees color symmetry, and reduces the number of parameters (as compared with the same network architecture without color symmetry) by $\sim q^2$, as in Eq.~\eqref{eq:PARsym}.

\subsection{MADE}

A Masked Autoencoder for Distribution Estimator (MADE) consists of a generic $D$-layer neural network satisfying the autoregressive property, i.e. the weights satisfy $J^{l}_{i\le j}=0$ for any layer $l=1,\cdots,D$.
When the depth $D=1$, we obtain the shallow model described in Sec.~\ref{sec:shallow}.
When $D>1$, we use ReLU activation functions between all layers. We can also increase the width of the hidden layers by defining a parameter $W$ that increases the dimensionality of each variable, i.e. $\left\{ x_1, \cdots, x_N \right\} \in \mathbb{R}^N \xrightarrow{} \left\{y_1^1, \cdots,y_1^{W}, y_2^1\cdots, y_N^{W}\right\} \in \mathbb{R}^{N\times W}$. Notice that when $W>1$, each variable $y_i^{a}$ satisfies the autoregressive property of depending only on $\{x_j\}_{j<i}$. To describe non-binary variables with the MADE architecture, we use the symmetric one-hot-encoding introduced in Sec.~\ref{sec:csym}.

\subsection{NADE}
\label{sec:NADE}

A Neural Autoregressive Distribution Estimator (NADE)~\cite{Uria2016} is another estimator based on an encoding-decoding architecture, where the autoregressive property is overimposed. It usually consists of a single hidden layer to encode the message, and the size of this hidden layer corresponds to $N_h$ hidden variables for each visible variable. This is similar to a MADE with $D=2$ and $W=N_h$. The $i=1,\cdots,N$ hidden units of a NADE are calculated from
\begin{equation}
    \vec{y}_i = \Psi\left[A \s_{<i} + \vec{B}\right],
\end{equation}
where $\vec{y}_i \in \mathbb{R}^{N_h}$, $A\in\mathbb{R}^{N_h\times N}$, $\s_{<i}$ is the usual $N$-dimensional input (each of the one-hot-encodings is processed in parallel to enforce color symmetry, see Sec.~\ref{sec:csym}) with the autoregressive mask, $\vec{B} \in \mathbb{R}^{N_h}$, and $\Psi$ is an arbitrary component-wise non-linear function, e.g. ReLu.
The distinctive feature of the NADE architecture is that the weight matrix $A$ and biases $\vec{B}$ are shared between all the hidden units $\vec{y}_i$, while in a MADE there would be a different $A_i$, $\vec{B}_i$ for each hidden unit, which can be encoded in fully connected weight matrices $\tilde{A} \in  \mathbb{R}^{N \times N_h \times N}$ and
$\tilde{B}\in \mathbb{R}^{N\times N_h}$.
This feature, inspired by restricted Boltzmann machines, has proven to be very convenient to reduce the number of parameters (by a factor $N$) while keeping a good expressivity of the model.
The information is propagated from the hidden layers to the output through
\begin{equation}
    H_i= \Psi\left[V_i\vec{y}_i + C_i\right],
\end{equation}
where $V_i \in \mathbb{R}^{N_h}$ and $C_i \in \mathbb{R}$ assume different values from each output field $i=1,\cdots, N$. Finally, a softmax is applied to the $q$ fields computed in parallel to obtain the probability, see Sec.~\ref{sec:csym}.

\subsection{ColoredMADE and ColoredNADE}

These architectures are modifications of the MADE and NADE architectures in order to describe non-binary variables in the standard one-hot-encoding fashion. In practice, this is realized by setting $W=q$, where $q$ is the number of colors, and one-hot-encoding the input. This choice does not enforce the color permutation symmetry, so each weight matrix is now free to take all possible values. This allows for more flexibility, at the cost of an increased memory consumption and training time.

\subsection{GADE}

A natural way to improve the performance of machine-learning models is to incorporate the physics of the problem into the model architecture. In the case of our spin models, the fact that the interactions have a graph structure suggests that Graph Neural Networks (GNN) may be an efficient choice. For this reason, we considered a GNN model that we called GADE, which stands for Graph Autoregressive Distribution Estimator. 

Following the message passing paradigm, see e.g.~\cite{pmlr-v70-gilmer17a}, we define $m_{i\xrightarrow[]{}j}$ as the message from node~$i$ to node~$j$. Each node has to perform the following operations: (i)~define the message to pass, (ii)~take the information from the incoming messages, (iii)~update its state.
Hence, for each of the $t=1,\cdots,T$ graph propagations we do the following operations:
\begin{equation}
   \mathrm{(i):}\;\; m^t_{i\xrightarrow[]{}j} = \begin{cases} \delta_{\s_i , \s_j} \cdot y^{t-1}_i \ ,& \mathrm{if }~i < j  \ , \\ 
    0~,& \mathrm{otherwise} \ .
    \end{cases}
    \label{eq:message}
\end{equation}
This operation imposes the autoregressive property by stating that a message goes from $i$ to $j$ only in the autoregressive direction. The message itself corresponds to the state of the node at the previous iteration, $y^{t-1}_i$, but it vanishes if the nodes have different colors. Notice that $y^0_i=\s_i$.
Then,  
\begin{equation}
    \mathrm{(ii)+(iii):}\;\; y^t_j = \text{Average}\left\{ m^{t}_{i\xrightarrow[]{}j} \; ,\, \forall i \in \partial j \right\} \ ,
    \label{eq:message2}
\end{equation}
where $\partial j$ are the nearest neighbors of $j$ on the graph.
Compared to commonly used GNN like GraphSAGE~\cite{NIPS2017_5dd9db5e}, the approach defined by Eqs.~(\ref{eq:message}-\ref{eq:message2}) does not require the introduction of any new model parameters. As such, it can be interpreted as a deterministic pre-processing operation that we add to the model pipeline.

After we have computed the graph-weighted state of each node, $\left\{ y^t_i \right\}_{t=0}^T$, we feed it to a MADE model (GADE) or to a ColoredMADE model (ColoredGADE). Note that since we have explored the graph up to a depth $T$, the MADE network requires $T$ input channels to process the data. This means that overall a GADE model will have the same number of parameters as a MADE with width $W=T$.  

\subsection{Variable ordering}

When constructing AR models using the
Bayes' rule decomposition,
\begin{equation}
    P_a(\s) = P_a^1(\s_1) P_a^2(\s_2|\s_1) \cdots
    P_a^N(\s_N | \s_{N-1},\cdots, \s_1) \ ,
    \nonumber
\end{equation}
the choice of the ordering of variables is arbitrary. In our case, e.g. for the random graph coloring problem, we can arbitrary label the variables as $\sigma_1,\cdots,\sigma_N$ and then construct the random graph. We call this the `naive' ordering. One can also consider smarter orderings~\cite{trinquier2021efficient}, in which variables are ordered according to some useful characteristics. In our work,
we considered a seemingly smarter ordering, in which the interaction graph is constructed first, and the variable $\sigma_1$ is taken to be the most connected one, $\sigma_2$ as the second most connected, and so on. The rationale for this choice~\cite{trinquier2021efficient} is that it can be useful to put more constrained variables first in the ordering.
Yet, we found that the choice of ordering does not affect the conclusions qualitatively, we thus stick to the naive ordering when not told otherwise.

\section{Adaptive training scheme}
\label{sec:adaptive}

The idea of this scheme is to perform a mixture of local and global MCMC, now at fixed temperature, starting with a decent guess for $P_{AR}(\s)$. The algorithm works in three steps:
\begin{enumerate}
    \item Perform a global MCMC step using a proposed $\s_{new}$ from $P_{AR}$.
    \item Perform a certain number of local MCMC steps.
    \item Repeat the first two steps $M$ times to generate a sample $\{\s^m\}_{m=1,\cdots, M}$. Use this sample to perform a few steps of gradient ascent on the log-likelihood to improve $P_{AR}(\s)$.
\end{enumerate}
Initially, most global MCMC moves are rejected, hence the sample is only constructed using local MCMC, resulting in poor sampling if local MCMC is inefficient. Yet, the little information contained in the sample might be enough to improve the AR model, resulting in a better acceptance of the global MCMC at the next step, and in turn of a sample of better quality. Ideally, the process should converge to a situation where the acceptance rate of global MCMC is high enough, thus achieving proper sampling; yet, this is far from being guaranteed in practical applications.

Furthermore, there is a danger of mode-collapse during the training~\cite{gabrie2021adaptive}. Suppose that global MCMC is very inefficient and that local MCMC is stuck into a probability peak of the true model. Then, at the first iteration, the sample only covers this single peak; the AR model will learn that peak, resulting in its reinforcement, and the process thus remains stuck. To prevent this, Ref.~\cite{gabrie2021adaptive} uses many parallel MCMC, initialized in very different states. Yet, in systems with a really large number of probability peaks (i.e., a rough energy landscape), this might not be enough to prevent mode-collapse.

\section{Details on the Curie-Weiss model}
\label{sec:CWdetails}

\subsection{The model and its Boltzmann distribution}

The model Hamiltonian is
\begin{equation}
    H = -\frac{1}{N} \sum _{i<j} \s_i \s_j = - \frac{N}{2} m^2 \ , \quad m = \frac{1}N \sum_i \s_i \ ,
\end{equation}
where an irrelevant constant of order one has been neglected in the second expression of $H$. Note that because all spin pairs interact, in order to have an extensive energy, the spin coupling must be fixed to $1/N$. Also, because the Hamiltonian is fully invariant under permutations of the spin labels, the ordering of variables in the AR model is irrelevant.

Because the number of spin configurations corresponding to a given $m$ is $\binom{N}{N (1+m)/2}$,
the partition function can be written as: 
\begin{equation}\label{eq:CWN}
    Z_N = \sum_m \binom{N}{N (1+m)/2} e^{\frac{\beta N m^2}2} = \sum_m e^{-N \beta f(m,T)} \ ,
\end{equation}
and the free energy per spin follows as
${f(T) = -(T/N) \log Z_N}$.
Using Stirling's formula to approximate the binomial coefficient for large $N$, one obtains
\begin{equation}\label{eq:CWNio}
\begin{split}
    &f(m,T) = - \frac{m^2}{2}  - T s(m) \ , \\ 
    &s(m) = -\frac{1+m}{2} \log \frac{1+m}{2} - \frac{1-m}{2} \log \frac{1-m}{2} \ .
\end{split}\end{equation}
In the thermodynamic limit, the value of $m$ that dominates the sum in Eq.~\eqref{eq:CWN} is the one that minimizes $f(m)$, which is the solution of $m = \tanh(\beta m)$, leading to $f(T) = \min_m f(m,T)$.

\subsection{Minimization of the KL divergence between the AR model and the mixture model}

We detail here the minimization of the KL divergence between the shallow MADE model,
\begin{equation}\nonumber
P^i_{AR}(\s_i|\s_{<i}) = \frac{\exp(\sum_{j(<i)} J_i \s_i\s_j)}{2\cosh ( \sum_{j(<i)} J_i \s_j )}
=\frac{\exp(J_i \s_i m_{<i})}{2\cosh ( J_i m_{<i} )}
\ ,
\end{equation}
where $m_{<i} = \sum_{j(<i)} \s_j$ is the magnetization of the first $(i-1)$ spins, and the mixture model that approximates the Boltzmann distribution of the CW model. 

For an infinite sample ($M\to\infty$, neglecting constants), the KL divergence between these two distributions is
\begin{equation}\label{eq:DJCW}
\begin{split}
    \mathcal{D}(\{J_i\}) 
    &= - \sum_i \langle \log P^i_{AR}(\s_i|\s_{<i}) \rangle\\
    &=-\sum_i \langle J_i \s_i m_{<i} -\log 2\cosh(J_i m_{<i}) \rangle
    \ ,
\end{split}
\end{equation}
where the average is over the mixture model
\begin{equation}
\begin{split}
P_B(\s) &\sim \frac{1}2 \prod_i p_+(\s_i) + \frac{1}2 \prod_i p_-(\s_i) \ , \\
p_\pm(\s_i) &= \frac{1 \pm \hat m \s_i}{2} \ .
\end{split}\end{equation}
Minimization of each term with respect to $J_i$
leads to a set of coupled equations:
\begin{equation}
     \langle \s_i m_{<i}-\tanh(J_i m_{<i}) m_{<i} \rangle = 0 \ .
     \end{equation}
     Because of the spin-flip symmetry, the average $\langle \s_i \s_j \rangle$ is the same in the two terms with $m=\pm \hat m$, , hence we can restrict to e.g. $m=+\hat m$ and write $\langle \s_i \s_j \rangle = \langle \s_i \s_j \rangle_+ = \langle \s_i \rangle_+ \langle \s_j \rangle_+ = \hat m^2$, which leads to
\begin{equation}
      (i-1)\hat m^2  = \langle \tanh(J_i m_{<i})m_{<i} \rangle \ .
\end{equation}
Similarly,
we can restrict to $m=+\hat m$ and write $m_{<i} = 2k-(i-1)$, $k$ being the number of positive spins among the first $(i-1)$. The probability distribution of $k$ is
a binomial,
hence we arrive to
\begin{equation}
    \begin{split}
    p(k)&={i-1\choose k}\left(\frac{1+\hat m}{2}\right)^k \left(\frac{1-\hat m}{2}\right)^{i-1-k} \ ,  \\
        \hat m^2 &= \sum_{k = 0}^{i-1} p(k) \frac{2k-i+1}{i-1}\tanh[J_i(2k-i+1)] \ .
    \end{split}
\end{equation}
The minimal value of the KL divergence is then
\begin{equation}
\begin{split}
    \hat{\mathcal{D}}& 
    = -N s(\hat m) - \sum_i J_i (i-1)\hat m^2\\
    &+\sum_i \sum_{k = 0}^{i-1} p(k) \log[2\cosh(J_i(2k-i+1))] \ .
\end{split}
\end{equation}

We can easily check that there is no problem in the convergence of the gradient descent algorithm in the vicinity of the phase transition, by expanding each of the terms $\mathcal{D}(J_i)=-\langle \log P^i_{AR}(\s_i|\s_{<i}) \rangle$ in Eq.~\eqref{eq:DJCW} for small $J_i$, which leads to
\begin{equation}
\begin{split}
\mathcal{D}(J_i) &= - A J_i + B \frac{J_i^2}2 + O(J_i^3) \ , \\
A&=(i-1) \hat m^2 \ , \\
B&=(i-1) + (i-1)(i-2) \hat m^2 \ .
\end{split}\end{equation}
For $\hat m=0$, this has a single minimum in $J_i=0$ with finite curvature $B=(i-1)$, leading to exponentially fast convergence. Even for $T<1$, the minimum is slightly shifted to a finite $J_i$ but its curvature stays finite.

\section{Additional results on the Edwards-Anderson model}
\label{sec:EAdetails}

\begin{figure*}[t]
  \centering
\begin{minipage}{.65\textwidth}
  \includegraphics[width=\textwidth]{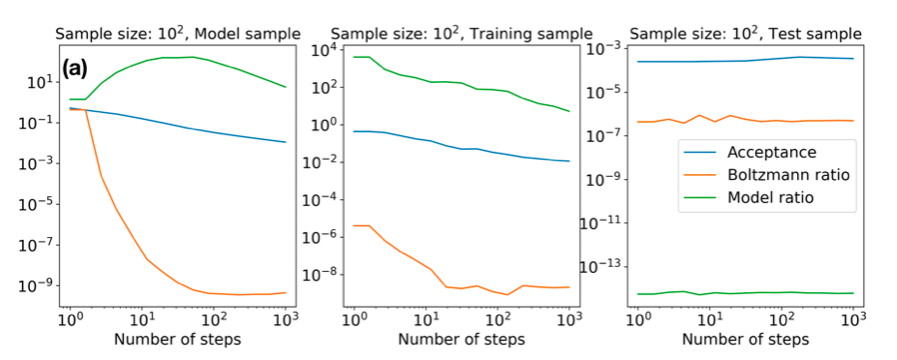}  
    \includegraphics[width=\textwidth]{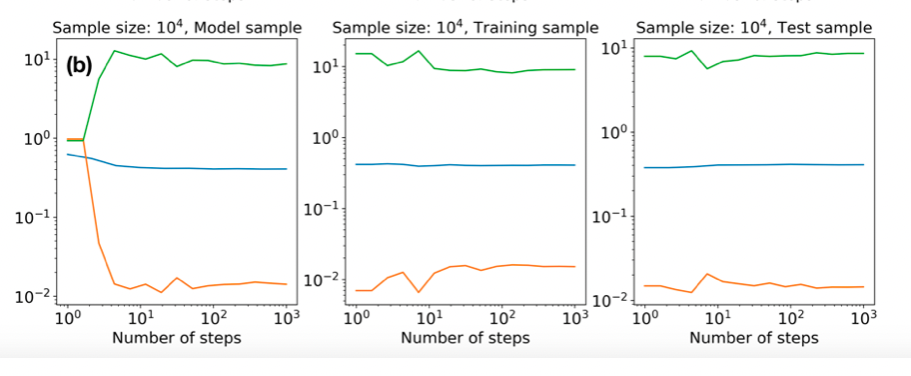} 
    \end{minipage}
    \begin{minipage}{.34\textwidth}
  \includegraphics[width=1.2\textwidth]{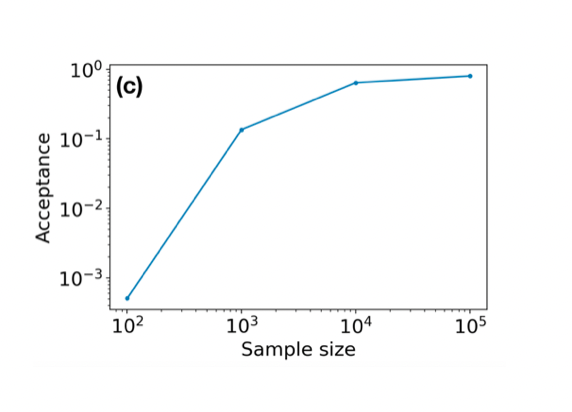} 
  \end{minipage}
    \caption{{\bf 2d Edwards-Anderson model} with $N=100$ spins. 
  (a)~Acceptance rate of global MCMC as a function of the number of MC steps, using a shallow MADE trained via maximum likelihood at $T=1$ on a sample of $M=10^2$ equilibrium configurations.
Here, $K=10^3$ independent global MCMC are initialized in configurations generated independently by the AR model itself (model sample), taken from the training sample, or from an independent equilibrium sample never seen by the model (test sample).
(b)~Same with $M=10^4$.
(c)~Acceptance rate of global MCMC (initialized on a test sample) as a function of training sample size $M$ for $T=1$.
}  
  \label{acceptance_EA}
\end{figure*}

\subsection{Dependence on the size of the training set}

In order to investigate the dependence on the size of the training set $M$,
we trained the NADE at fixed temperature $T=1$ by using a smaller training sample of $M=10^2$ (obtained by subsampling the original sample of $M=10^5$ configurations).
We then studied the acceptance rate of global MCMC as a function of the number of MC iterations. We consider $K=10^3$ independent chains, initialized in three different ways: (i) using samples generated by the NADE itself; (ii) using the samples in the training set; and (iii) using independent equilibrium samples obtained by local MCMC.
One should keep in mind that, because
 the test sample initialization~(iii) correspond to an initial sample in equilibrium and global MCMC satisfies detailed balance, the system remains in equilibrium at all times, and all quantities must then be time-independent, which is confirmed in Fig.~\ref{acceptance_EA}a. 
 Also, this initialization thus corresponds to the long-time limit of the other two, (i) and~(ii).
In Fig.~\ref{acceptance_EA}a we show, as a function of the number of global MC steps, the acceptance rate averaged over the $K$ chains, together with the geometric average over the $K$ chains of 
the Boltzmann ratio $P_B(\s_{new})/P_B(\s_{old})$ and the AR model ratio $P_{AR}(\s_{old})/P_{AR}(\s_{new})$.
We recall that, according to Eq.~\eqref{eq:ARglobal}, the acceptance rate depends on the product of the Boltzmann and model ratios.
 We observe that at $M=10^2$ the asymptotic acceptance rate (as obtained from the test set initialization) is poor, of the order of $10^{-3}$.
Furthermore, the Boltzmann and model ratios are both much smaller than one, but have very wide fluctuations in log-scale, hence the rare acceptance of moves is due to an atypical fluctuation of these ratios. When the global MCMC is instead initialized in the model or training samples, the acceptance rate is initially much higher and decreases very slowly, indicating that the global MCMC is behaving differently on these samples, and that it is taking an extremely long time to reach the asymptotic behavior from these initializations. We conclude that with a training set of $M=10^{2}$ samples, the global MCMC has poor decorrelation properties, and is generally inefficient, because the AR model is overfitting the training set.
We then repeat the same study for larger values of $M$, and we find that around $M=10^4$ the acceptance rate becomes good (Fig.~\ref{acceptance_EA}b), and correspondingly the dependence on initialization disappears. This justifies the choice of $M=10^5$ as an optimal value, and highlights the need for a sufficiently large training set to obtain a good performance of global MCMC. The dependence of the acceptance rate on $M$ is reported in Fig.~\ref{acceptance_EA}c.

\begin{figure*}[t]
  \centering
  \includegraphics[width=\textwidth]{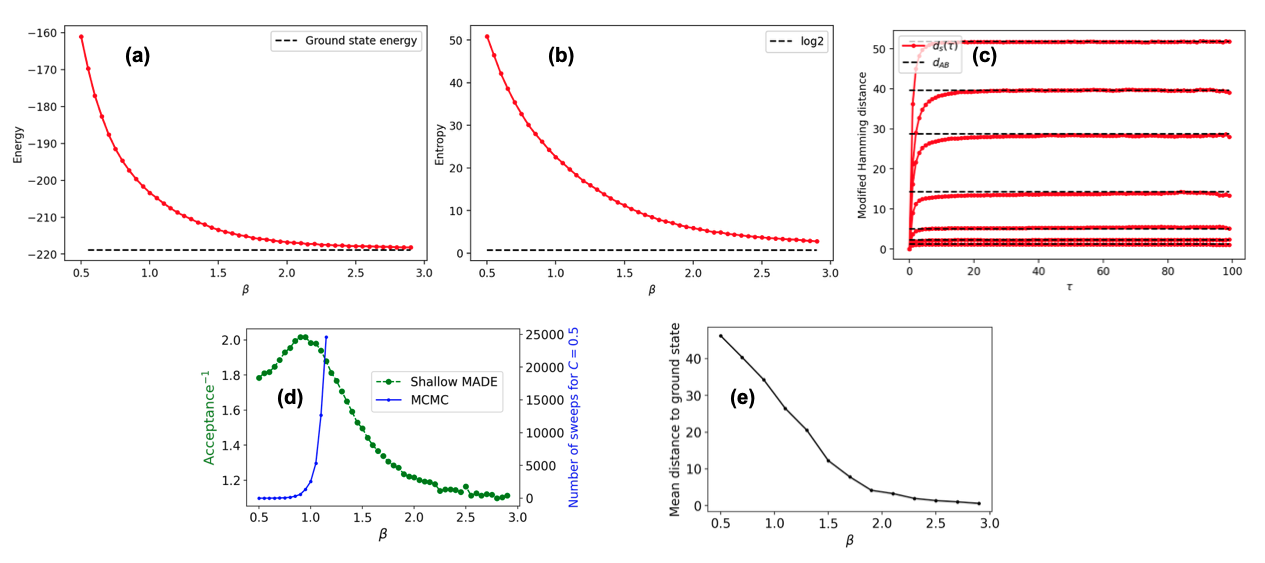}  
    \caption{{\bf 3d Edwards-Anderson model} with $N=125$ spins ($5\times 5 \times 5$ periodic cubic lattice), using as AR model a shallow MADE trained via sequential tempering with a training sample of $M=10^5$ configurations. (a)~Total energy and (b)~total entropy as a function of $\beta$, as estimated by the shallow MADE, compared with the exact ground state energy obtained via the McGroundstate server~\cite{CJMM22} and the corresponding ground state entropy ($\log 2$).
(c)~Average Hamming distance $d_{AB}$ between two independent global MCMC runs (dashed horizontal lines), compared with the average Hamming distance $d_s(\tau)$ between a single global MC chain at two distinct times separated by $\tau$ global MC steps (red dots and lines). From top to bottom, $\beta=0.5, 0.9, 1.3, 1.7, 2.1, 2.5, 2.9$.
(d)~Inverse of the acceptance rate of global MCMC as a function of inverse temperature $\beta$, compared to the decorrelation time of local MCMC. The latter diverges around the critical temperature $\beta\sim 1$, where the former shows a local maximum.
(e)~Average Hamming distance (i.e. number of flipped spins) between the ground state and equilibrium configurations generated by global MCMC, as a function of $\beta$.
}  
\label{SI:fig3d}
\end{figure*}

\subsection{Three-dimensional model}

We also investigated the three-dimensional Edwards-Anderson (3d EA) model, on a $5\times5\times5$ cubic periodic lattice, i.e. with $N=125$ spins. Contrarily to the 2d case, finding the exact ground state in the 3d case is NP-hard~\cite{barahona1982computational}.
Yet, for these small sizes, 
the McGroundstate server~\cite{CJMM22} can find the ground state in a few minutes.
Furthermore, in the 3d case it is known that a phase transition happens at a critical temperature very close to $T=1$~\cite{KKY06}.

We find that global MCMC using the shallow MADE architecture and a training set of $M=10^5$ samples performs very similar to the 2d case. In particular, it provides a diverging speedup with respect of local MCMC (whose decorrelation time diverges near the phase transition) and converges to the ground state at low temperatures. The results are reported in Fig.~\ref{SI:fig3d}.

\end{document}